\documentclass[useAMS,usegraphicx,usenatbib]{mn2e}
\newcommand\msun{\rm{M_{\odot}}}

\def\stacksymbols #1#2#3#4{\def\theguybelow{#2}
        \def\verticalposition{\lower#3pt}
        \def\spacingwithinsymbol{\baselineskip0pt\lineskip#4pt}
        \mathrel{\mathpalette\intermediary#1}}
\def\intermediary #1#2{\verticalposition\vbox{\spacingwithinsymbol
        \everycr={}\tabskip0pt
        \halign{$\mathsurround0pt#1\hfil##\hfil$\crcr#2\crcr
                \theguybelow\crcr}}}

\def\lta{\stacksymbols{<}{\sim}{2.5}{.2}}
\def\gta{\stacksymbols{>}{\sim}{2.5}{.2}}

\title[AGN Outflows: Regulating the Dance of Heating and Cooling]
{The Dance of Heating and Cooling in Galaxy Clusters: \\ 
3D Simulations of Self-Regulated AGN Outflows}

\author[M. Gaspari et al.]{M. Gaspari$^{1}$\thanks{E-mail:
massimo.gaspari4@unibo.it}, C. Melioli$^1$,
F. Brighenti$^1$, A. D'Ercole$^2$ \\
$^{1}$Astronomy Department, University of Bologna, Via Ranzani 1, 
40127 Bologna, Italy\\
$^{2}$INAF-OABO, Via Ranzani 1, 40127 Bologna, Italy}
\begin{document}

\date{Accepted 2010 September 10.  Received 2010 August 31; in original form 2010 June 2}

\pagerange{\pageref{firstpage}--\pageref{lastpage}} \pubyear{2010}

\maketitle

\label{firstpage}

\begin{abstract}
It is now widely accepted that heating processes play
a fundamental role in galaxy clusters, struggling in an intricate but 
fascinating `dance' with its antagonist, radiative cooling. Last 
generation observations, especially X-ray, are giving us tiny hints about
the notes of this endless ballet. Cavities, shocks, turbulence 
and wide absorption-lines indicate the central active nucleus
is injecting huge amount of energy in the intracluster medium.
However, which is the real dominant engine of self-regulated heating?\\
One of the model
we propose are massive subrelativistic outflows, probably generated by a wind disc or
just the result of the entrainment on kpc scale by the fast radio jet.
Using a modified version of AMR code FLASH 3.2, we explored several feedback mechanisms 
which self-regulate the mechanical power. Two are the best schemes that answer our primary
question, id est quenching cooling flow and at the same time preserving a cool core
appearance for a long term evolution (7 Gyr): one more explosive (with efficiencies $\sim5\times10^{-3}
-10^{-2}$), 
triggered by central cooled gas, and
the other gentler, ignited by hot gas Bondi accretion (with $\epsilon=0.1$). 
These three-dimensional simulations show that the total energy injected is not
the key aspect, but the results strongly depend on how energy is given to the ICM.
We follow the dynamics of best models (temperature, density, SB maps and profiles) and
produce many observable predictions:
buoyant bubbles, ripples, turbulence, iron abundance maps and hydrostatic equilibrium
deviation. We present a deep discussion of merits and flaws of all our models, 
with a critical eye towards observational concordance.

\end{abstract}


\section[]{Introduction}

A fundamental gap in our understanding of the formation and evolution
of galaxies and galaxy clusters concerns the thermal evolution of the 
baryonic component of these systems (e.g. \citealt{mcn07} [MN07];
\citealt{cat09}).
Massive dark matter halos contain
large amount of hot gas, shining in the X-ray band. The observed radiative 
losses, if not compensated for by some kind of heating, would imply
gas cooling rates ranging from $\sim 1$ M$_\odot$ yr$^{-1}$, for
massive elliptical galaxies, to hundreds M$_\odot$ yr$^{-1}$ for rich 
clusters (\citealt{fab94,per98}). However, since the first XMM-RGS observations
it has been clear that the (radiative) cooling rate in clusters and galaxies 
is reduced by at least one order of magnitude with respect to the simple expectation
(\citealt{pet01,pet03,xue02,pef06}, and references therein). 
This is the so-called `cooling flow problem'.

Active galactic nuclei (AGN) can easily provide enough energy to the gas to offset the energy lost
by radiation and high resolution X-ray images show indeed clear evidence of  
AGN-gas interaction in many clusters and galaxies
(\citealt{boe93,bla01,fij01,jon02}; MN07 and references therein). 
The fairly common presence of X-ray cavities, 
often coincident with lobes of radio emission
connected to the core of the central galaxy by a radio jet, indicates 
that AGN inject energy in the intracluster medium (ICM) 
in kinetic form (outflows) and as relativistic particles, 
although the quantitative significance of the latter
is difficult to estimate (\citealt{dun05,git09}).

\citet{raf06} and \citet{raf08} showed that the AGN power
associated to the cavity formation is of the same order 
as the core X-ray luminosity, for a sample of 33 clusters and groups.
Although this energetic balance
is only a necessary but non sufficient requirement for a
heating scenario to be successful, it strongly suggests
that the heating process manifests itself
generating bubbles in the ICM.
Cavities can be easily created by `directional' input of energy,
such as jets or collimated outflows,
making spherically symmetric form of heating less appealing
as major players in solving the cooling flow problem.
 
In order to prevent significant gas cooling, the feedback process must be activated
with a frequency not greatly different
from $1/t_{\rm cool}$, where $t_{\rm cool}$ is the central cooling time, often of the
order of ${\rm few}\;\times 10^8$ yr for clusters (e.g., \citealt{san06,mit09})
and even lower for elliptical galaxies or groups
(e.g. $t_{\rm cool}\sim 1.5\times 10^7$ yr for NGC 4636, see \citealt{bal09}).

Moreover, the feedback heating
must preserve the cool core appearance of the
majority of the clusters (\citealt{per98,mit09}).
In fact, it has been shown
that concentrated heating while very efficient in stopping the cooling
process, often generates negative temperature gradients, contrary to 
the observations (\citealt{brm02,brm03,mat06}).
Another indication that AGN cannot deposit most of its energy in the very
central region is the common survival of galactic scale cool cores
in cluster ellipticals (\citealt{sue05,sue07}); a spatially concentrated
heating would easily erase these fragile low temperature regions (\citealt{brm02,brm03}).

Motivated by these considerations,
we investigate here the long term effect of kinetic feedback on the
ICM (see MN07 and \citealt{fab09}). 
We assume that AGN outbursts generate collimated, subrelativistic
outflows on kpc scale. There is widespread observational evidence 
for winds originating in galactic nuclei. High redshift radio
galaxies host galactic scale, bipolar outflows of ionized gas
with velocity $\sim 1000$ km s$^{-1}$, likely triggered
by the interaction of the radio jet with the ISM (\citealt{nes08}).
Morganti et al. (2005, 2007) report the detection of fast massive
neutral outflows, using 21-cm blueshifted absorption lines against 
strong radio continuum. They occur at
kpc distance from the nucleus with rates of tens M$_\odot$ yr$^{-1}$.
Optical, UV and X-ray observations of highly ionized
gas also point toward fast outflows (thousands km s$^{-1}$) driven by
entrainment, even if the mass rates are relatively modest
(\citealt{geo98,cre99,kri03,ris05,mck07,por09}; see \citealt{cre03} for a review).
It is reasonable to conclude that most of the AGN exhibit outflows,
albeit the physical parameters of the winds are still uncertain.
The geometry of the flow is also unclear, both polar
winds, perhaps caused by entrainment of the ICM in the relativistic 
radio jet, or equatorial disc winds (see \citealt{pro07} and references
therein) being possible.

The key question we want to address
in this work is the following: {\it are outflows from the central AGN
able to prevent the ICM from cooling
and at the same time preserve the cool core appearance?}
 
In recent years a considerable amount of research has been 
devoted to the understanding of the effect of jets on the ICM.
Most works investigated the transient flow resulting from
a bipolar outflow
(\citealt{rey01,rey02,baa03,rus04,omm04,zan05,bru07,ste07,sts09}),
but did not fully confronted the question above.  

The long term influence of AGN outflows on cluster cooling flows 
was studied by \citeauthor{brm06} (2006, hereafter BM06).
They used 2D simulations 
and a mechanical feedback scheme, self-regulating injection time 
(but not velocity and power). They calculated the
gas cooling rate and the azimuthally averaged density and temperature
profiles.
It was found that some intermittent bipolar outflow models, with velocity in the
range $5\times 10^3 - 10^4$ km s$^{-1}$, could shut down
gas cooling for many Gyr, while preserving the cool core appearance
of the cluster. 

\citet{cat07} performed 3D calculations
of AGN feedback in a poor cluster, about $\sim 10$ time
less massive than the object studied here. They employed
a hybrid kinetic-thermal feedback which injects energy
at a rate proportional to the Bondi accretion rate to the
central black hole. With the adopted ICM initial conditions
the central cooling time is quite long (4 Gyr), and the average
cooling rate in absence of feedback is $\sim 30$ M$_\odot$ yr$^{-1}$.
During the most powerful AGN outbursts only a small fraction
of energy is ejected in kinetic form, the velocity of the jets
being only $\sim 1.4 \times 10^3$ km s$^{-1}$.
Their models are successful in stopping gas cooling,
but the cool core disappear after a few Gyr.
\citet{dub10} implemented this Bondi feedback in a cosmological
context, with similar results.

The present paper builds on the preliminary results by BM06 to explore
a larger set of feedback schemes and bound
the parameter space for successful feedback models.
In this work we use
multigrid, 3D hydrodynamical simulations for two reasons. First,
the nature of cooling flows is intrinsically chaotic and turbulent, therefore 
it is essential to allow a realistic description of all
instabilities, which in turn influence the outflow evolution. 
Second, we want also to eliminate the spurious cooling of gas along the $z$ 
symmetry axis present in 2D axisymmetric calculations
(like in BM06).

As stated above, our primary objective here is to investigate if
massive, collimated outflows can provide
a suitable form of feedback to solve the cooling flow problem. 
The adopted scheme for the generation of
jets and their connection with the accretion on the black hole
is simplified and parametrized (as customary in similar studies),
and attempt to investigate subjects like 
growth of the black hole or jet physics is certainly superficial.
We will critically test our models, trying to recognize
all the positive and problematic aspects of this form of feedback, comparing our
results with a great variety of observational constraints.

Finally, we mention that other authors simulated the feedback 
from buoyant cavities and the
associated shock heating, without explicitly including the jets that likely
generate these features in real clusters (\citealt{brk02,brm03,rus04,dal04,bru05,sij07,brs09,mab08a,mab08b,mat09}; 
the latter three papers considered cavities generated by cosmic rays).
This and other different kinds of feedback will be also explored by us
in forthcoming papers.

\section[]{The Computational Procedure}

The simulations presented here were calculated
with an highly modified version of FLASH 3.2
(\citealt{fry00}), a 3D adaptive mesh refinement (AMR) public code, 
which solves the hydrodynamic euler equations through a split Piecewise-Parabolic Method (PPM) 
solver, particularly appropriate to describe shock fronts. It uses the Message-Passing Interface 
(MPI) library to achieve portability and efficient scalability on a variety of different 
parallel HPC systems. The simulations were run on 128 processors of IBM P575 Power 6 (SP6) 
at CINECA supercomputing centre.

We added several source and sink terms to the usual hydro-equations solved by FLASH, in conservative form:
\begin{equation}\label{cont}
\frac{\partial\rho}{\partial t} + \bmath{\nabla}\cdot\left(\rho \bmath{v}\right) = \alpha \rho_{\ast} -q\frac{\rho}{t_{\rm {cool}}} + S_{1,\rm{jet}}\:,
\end{equation}
\begin{equation}\label{mom}
\frac{\partial\rho \bmath{v}}{\partial t} + \bmath{\nabla}\cdot\left(\rho \bmath{v} \otimes \bmath{v}\right) + \bmath{\nabla}{P} = \rho\bmath{g}_{\rm {DM}} + S_{2,\rm{jet}}\:,
\end{equation}
\[
\frac{\partial\rho \varepsilon}{\partial t} + \bmath{\nabla}\cdot\left[\left(\rho \varepsilon + P\right) \bmath{v}\right] = \rho\bmath{v}\cdot\bmath{g}_{\rm DM} + \alpha \rho_{\ast}\left(\varepsilon_0 +\frac{\bmath{v}^2}{2}\right) \nonumber
\]
\begin{equation}\label{ene}
\quad\quad\quad\quad\quad\quad\quad\quad\quad\quad\quad -\: n_{\rm{e}} n_{\rm{i}} \Lambda(T,Z) + S_{3,\rm{jet}}\:,
\end{equation}
\begin{equation}\label{eos}
P = \left(\gamma -1\right)\rho \left(\varepsilon-\frac{\bmath{v}^2}{2}\right)
\end{equation}
where $\rho$ is the gas density, $\bmath{v}$ the velocity, $\varepsilon$ the specific 
total energy (internal and kinetic), $P$ the pressure, $\bmath{g}_{\rm {DM}}$ the gravity of dark matter, and 
$\gamma = 5/3$ the adiabatic index. 
The temperature is computed from $P$ and $\rho$ using state Eq. (\ref{eos}), with an atomic weight $\mu\simeq 0.62$, appropriate for a totally ionized plasma with 25\% He in mass.

Outflow source terms $S_{1,2,3,\rm{jet}}$ (dependent on injected density, momentum
and mechanical energy, respectively),
along with the spatial distribution of the source region,
will be explained for every type of feedback in Section 2.2. Note that injection will be done
directly into the domain (without a mass inflow, $S_1 = 0$) or through boundary condition at $z=0$ ($S_1 > 0$). 

Radiative cooling is treated in the code like a source term in Eq. (\ref{ene}): the ICM loses 
energy at a volume rate $n_{\rm {e}} n_{\rm {i}} \Lambda(T, Z)$,
where $n_{\rm {e}}$ and $n_{\rm {i}}$ are the number density of electrons and ions, 
and $\Lambda$ is the cooling function for a temperature $T$ and metallicity $Z$
(\citealt{sud93}). In the following
we assume $Z = 0.3$ Z$_\odot$, 
a typical metallicity for these systems (\citealt{tam04}),
ignoring central negative gradients. The minimum temperature allowed is
$10^4$ K.

We also consider SNIa and stellar winds heating in the central
elliptical galaxy ($\rho_{\ast}$ is the de Vaucouleurs stellar density profile), 
although their effect is minor in massive clusters.
They are implemented following the parametrization of \citet{brm02}, 
with a rate (dominated by stellar mass loss)  
$\alpha = \alpha_{\ast} + \alpha_{\rm {SN}} \sim 4.7 \times 10^{-20} (t/t_{\rm{n}})^{-1.3}$ s$^{-1}$, where
$t_{\rm{n}} = 13.7$ Gyr is the present time, and with a specific injection energy $\varepsilon_0$ dependent
on SNIa and stellar winds temperature. 

We also add a mass dropout term (see \citealt{brm02}) to avoid a
clutter of zones filled with cold gas, which jet events could spread at larger radii
or, without feedback, accumulate in the nucleus. In fact, without this term
the simulation will generate clouds of very cold gas, whose physics
(collapse, star formation)
cannot be described by the hydrocode.
Thus, in order to prevent this feature, we include 
a mass sink term $-q(T)\rho/t_{\rm{cool}}$ in Eq. (\ref{cont}) 
and drop out the cold gas (at constant pressure). Its effect is simply
to remove the cold gas from the grid, without 
affecting the hotter flow or the calculated cooling rate (\citealt{brm02}).
The dimensionless coefficient $q(T)$ is defined as $2 \exp (-(T/T_{\rm {q}})^2)$
and dropout becomes significant when $T \lta T_q = 5 \times 10^5$ K. 
The cooling time is assumed $t_{\rm{cool}}=5P/2n_{\rm {e}} n_{\rm {i}}\Lambda$.
Note that the total mass of cooled gas does not depend on the presence of the dropout term or its functional form (\citealt{brm00}).

We calculate the flow evolution for 7 Gyr (large cluster formed relatively recently),
as opposed to most of the works cited in Section 1. 
As shown in Section 3.1 several
feedback heating schemes are only able to delay excessive gas cooling for 1-2 Gyr
before failing, and therefore we stress the need to investigate 
the long term (several Gyr) behaviour of heated flows.

\subsection[]{The cluster model and initial conditions}

\begin{table*} \label{params}
\begin{minipage}{200mm}
\caption{Parameters and properties of all simulated models.}
\begin{tabular}{@{}lcccccc}
\hline 
       &                       &  $\epsilon$      & $W_{\rm jet}$   & $Z_{\rm jet}$          & $v_{\rm jet}$         & $\tau_{\rm jet}$      \\ 
Model  & Feedback              &  efficiency      & jet width (kpc) & jet height (kpc)       &  jet velocity (km s$^{-1}$)  & duration/cycle (Myr) \\ 
\hline
 CF    & no AGN heating        &      -           &   -        & -           &     -            &     -           \\
 A1    & $\Delta M_{\rm cool}$ & $5\times10^{-4}$ & $2.7$      & $6.8$       & $(2 \epsilon \Delta M_{\rm cool} c^2/ M_{\rm act})^{1/2}$      &          self-regulated     \\    
 A2    & $\Delta M_{\rm cool}$ & $10^{-3}$        & $2.7$      & $6.8$       & $(2 \epsilon \Delta M_{\rm cool} c^2/ M_{\rm act})^{1/2}$      &          self-regulated     \\
 A2L   & $\Delta M_{\rm cool}$ & $10^{-3}$        & $2.7$      & $17$        & $(2 \epsilon \Delta M_{\rm cool} c^2/ M_{\rm act})^{1/2}$      &          self-regulated     \\
 A3    & $\Delta M_{\rm cool}$ & $5\times10^{-3}$ & $2.7$      & $6.8$       & $(2 \epsilon \Delta M_{\rm cool} c^2/ M_{\rm act})^{1/2}$      &          self-regulated     \\
 A3L   & $\Delta M_{\rm cool}$ & $5\times10^{-3}$ & $2.7$      & $17$        & $(2 \epsilon \Delta M_{\rm cool} c^2/ M_{\rm act})^{1/2}$      &          self-regulated     \\      
 A3S   & $\dot{M}_{\rm cool}$  & $5\times10^{-3}$ & $2.7$      & $0$         & $(2\epsilon \Delta M_{\rm cool} c^2/
\rho_{\rm jet} A \Delta t)^{1/3}$ & self-regulated     \\
 A4    & $\Delta M_{\rm cool}$ & $10^{-2}$        & $2.7$      & $6.8$       & $(2 \epsilon \Delta M_{\rm cool} c^2/ M_{\rm act})^{1/2}$      &          self-regulated     \\
 A4L   & $\Delta M_{\rm cool}$ & $10^{-2}$        & $2.7$      & $17$        & $(2 \epsilon \Delta M_{\rm cool} c^2/ M_{\rm act})^{1/2}$      &          self-regulated     \\
 A5    & $\Delta M_{\rm cool}$ & $5\times10^{-2}$ & $2.7$      & $6.8$       & $(2 \epsilon \Delta M_{\rm cool} c^2/ M_{\rm act})^{1/2}$      &          self-regulated     \\
 B1    & intermittent          &    -             & $2.7$      & $6.8$       &   $10^4$         &       20/200     \\
 B2    & intermittent          &    -             & $2.7$      & $6.8$       &   $10^4$         &       10/100     \\
 B3    & intermittent          &    -             & $2.7$      & $6.8$       &   $10^4$         &        1/10      \\
 C1    & continuous            &    -             & $2.7$      & $6.8$       &   $2\times10^3$  &   continuous     \\
 C2    & continuous            &    -             & $2.7$      & $6.8$       &   $6\times10^3$  &   continuous     \\
 C3    & continuous            &    -             & $2.7$      & $6.8$       &   $10^4$         &   continuous     \\
 BONDI &    $\dot{M}_{\rm B, 10 kpc}$  &    $10^{-1}$     & $2.7$      & $6.8$       &   $(2 \epsilon \dot{M}_{\rm B} \Delta t c^2/ M_{\rm act})^{1/2}$ &          self-regulated     \\
 BONDI2&    $\dot{M}_{\rm B, 5 kpc}$   &    $10^{-1}$     & $2.7$      & $0$         &   $(2\epsilon \dot M_{\rm B} c^2/\rho_{\rm jet} A)^{1/3}$      &          self-regulated     \\
\hline

\end{tabular}
\end{minipage}
\end{table*}

As in BM06, we adopt the well observed cluster Abell 1795
(\citealt{tam01,ett02})
as a template for our models. Being A 1795 a rather typical,
relaxed (\citealt{but96}), cool core massive ($M_{\rm vir} \sim 10^{15}$
M$_\odot$) cluster,
all the results we present should be relevant for any object 
in this category.
The short central cooling time, $t_{\rm cool} \sim 4 \times 10^8$ yr
(\citealt{ett02}) assures that this cluster should host a strong
cooling flow in absence of appropriate feedback.
However, recent X-ray observations place an upper limit to
the radiative cooling rate
$\dot M_{\rm cool} \lta 30 $ M$_\odot$ yr$^{-1}$ (\citealt{pet03,bre06}), 
which is a key constraint for the success of a model.

We start our calculations with the hot gas in spherical hydrostatic equilibrium 
in the potential well of the dark matter halo:
\begin{equation}\label{EH}
\frac{dP}{dr}(r) = -\rho(r)\frac{d\phi_{\rm DM}}{dr}(r)\:,
\end{equation}
where $r$ is the spherical radius.
The dark matter halo follows
a NFW distribution (\citealt{nfw96}), with virial mass $10^{15}$ M$_\odot$, and thus a potential
given by: 
\begin{equation}\label{NFW}
\phi_{\rm DM}(r) = -\frac{G M_{\rm vir}} {r_{\rm s} f(c_{102})} \frac{{\rm ln}(1+r/r_{\rm s})}{r/r_{\rm s}} \:,
\end{equation}
where the concentration at overdensity 102 is $c_{102}=r_{\rm vir}/r_{\rm s}\simeq6.6$ with
virial radius $\simeq2.6$ Mpc, and
$f(c_{102})={\rm ln}(1+c_{102})- c_{102}/(1+ c_{102})$. 
We have adopted a $\Lambda$CDM cosmological universe with $\Omega_{\rm M} = 0.27$, $\Omega_{\rm \Lambda} = 0.73$ and
$H_0 = 71$ km s$^{-1}$ Mpc$^{-1}$.  

Combining Eq. (\ref{NFW}) and an observed $T(r)$ fit (see third panel of Fig. 1, dotted line) with Eq. (\ref{EH}),
we recover the density radial profile $\rho(r)$, assuming an ideal gas $P=k_{\rm b}\rho T/\mu m_{\rm p}$
and a gas fraction of 0.15 at virial radius..

In most models we ignore
the contribution to the gravitational potential due to the central galaxy. This is important
only in the inner few tens kpc and has no significant effect on the cooling rate or
the hydrodynamical variable profiles. In one model we have also included the gravity from the central galaxy,
modelled as a de Vaucouleurs profile (\citealt{mem87}) with total stellar mass $M_{\ast}$ $\sim 6 \times 10^{11}$ 
M$_\odot$ and effective radius $r_{\rm {e}}$ $\sim 8.5$ kpc, 
and verified that its influence on the results is inconsequential (see Section 3.1.3).

The computational rectangular box in all of our models extends slightly beyond the cluster
virial radius $r_{\rm vir}$. We simulate just half cluster with symmetric boundary
condition at $z=0$, while elsewhere we set prolonged initial conditions with only outflow permitted.
Despite the AMR capability of FLASH, we decided to use a number of concentric
{\it fixed} grids in cartesian coordinates. This ensures a proper resolution of 
the waves and cavities generated in the cluster core by the AGN outflows.
We use a set of 9 grid levels (with basic blocks of $8\times8\times4$ points),
with the zone linear size doubling among adjacent levels.
The finest, inner grid 
has a resolution of $\sim2.7$ kpc and covers a spherical region of 100 kpc in radius.
In general, grids of every level extend radially for about 40 cells.
The relatively low resolution is due to the need of covering large spatial scales
(kpc up to Mpc) and at the same time integrating the system for several Gyr,
using moderate computational resources (50,000 CPU hours).

\subsection[]{Outflow generation}
We adopt a purely mechanical AGN feedback in form of nonrelativistic,
collimated outflows (similar to BM06). In this paper we show results only for models
with cylindrical jets, with velocity
parallel to the $z-$axis. We have calculated few simulations with
conical outflows (with half-opening 
angle up to $70$ degrees -- see also \citealt{ste07})
and we have verified that they have a similar
impact on the global properties of the flow. In fact, the pressure of
the ICM collimates the outflows within few tens kpc
(see BM06).

We consider several types of feedback. In feedback scheme A
an outflow is activated only when gas cools to very low
temperature within a 
spherical region $r<10$ kpc, and drops out
from the flow (conceptually, this is similar to the ``cold
feedback model'' described by \citealt{pis05}).
Usually most of the gas cools
at the very centre of the cluster.
We assume that at any timestep a fraction $\epsilon$ of the rest mass energy of the 
cooled gas, $\Delta M_{\rm cool} c^2$, is injected as kinetic energy.
Here $\Delta M_{\rm cool}$ is the gas mass cooled in a given timestep of the finest
grid. This energy 
is given to the hot gas located in a small region at the centre of the grid
(the `active jet region'), whose size is indicated in Table 1 and
containing a gas mass $M_{\rm act}$ (there is no new injected mass in Eq. (\ref{cont})). 
At every timestep we set the $z$ component of the velocity within the active region to
$v_{\rm j} = (2 \epsilon \Delta M_{\rm cool} c^2/ M_{\rm act})^{1/2}$, since
$E_{\rm k,jet} \equiv 0.5M_{\rm act}v_{\rm jet}^2 = \epsilon \Delta M_{\rm cool} c^2$. 
We will see
(Section 3) that the frequency and strength of the feedback events
strongly depend on the mechanical efficiency $\epsilon$, which has typical values
$10^{-4} - 10^{-2}$.

We remark again that this scheme to link gas cooling, 
black hole accretion and outflow generation does not have a strong physical basis:
it must be taken as a simplified way to implement a self-regulated
feedback, which triggers heating only when it is needed to halt ICM cooling.
In considering massive slow outflows, we are implicitly assuming that the relativistic
radio jet entrains some ICM mass ($M_{\rm act}$). Moreover, other authors (e.g. \citealt{gio04}) 
found that radio jets in cluster central galaxies are highly relativistic on pc scale, 
but rapidly decrease to subrelativistic velocities within few kpc from the black hole (especially 
in Fanaroff-Riley I sources), 
because of the interaction with the dense ISM in the inner region.

In feedback method B the outflows are triggered intermittently, at fixed
times (Table 1) and
a fixed velocity $v_{\rm jet}$ is given to the gas
located in the active region (again $S_1=0$ in Eq. (\ref{cont})). Feedback scheme B is not self-regulated,
but the AGN outbursts are forced to occur with a frequency (typically
of the order of $10^7 - 10^8$ yr$^{-1}$) which agrees with the observational estimates
(\citealt{saf07,saf08}).
The power of any jet event depends only on the
mass present in the source region and is
independent of the accretion rate.

Next we run a few models where the outflows are continuously generated,
that is at every time step we keep the $z-$component of the velocity 
within the jet active region at a given value, listed in Table 1 (scheme C).

We also adopt a scheme, named BONDI, in which the accretion rate is calculated with
the Bondi prescription (see also \citealt{cat07}):
$\dot M_{\rm Bondi} = 4\pi (GM_{\rm  BH})^2\rho_0/c^3_{{\rm s}0}$, where $\rho_0$
is the volume-weighted hot gas density calculated within $r\lta 5 $ (or $10$) kpc,
while $c_{{\rm s}0}$ is the mass-weighted sound speed in the same region.
Needless to say, the Bondi radius, $r_{\rm B}\sim 50$ pc, is far smaller
than our resolution limit, so we refrain to attach a strict
physical meaning  to $\dot M_{\rm B}$. In this sense the high mechanical efficiencies ($0.1$) used for
Bondi models are due to the fact that the accretion should be a factor 10-100 larger (because
of higher inner $\rho_0$ and lower $c_{{\rm s}0}$). This feedback is fundamentally different
from feedback A because it produces a self-regulated, quasi-continuous, low power AGN
activity, in contrast to the few violent AGN outbursts characteristic
of the best models adopting scheme A.

Finally, in BONDI2 and A3S the outflow is not injected as usual in the active region, but with a mass, momentum and
energy flux through the boundary at $z=0$ (thus $S_1>0$), with a square area $-1.35 \lta x\lta 1.35$ kpc,
$-1.35 \lta y \lta 1.35$ kpc. 
The velocity of the jet is then calculated as $v_{\rm jet} = (2\epsilon \dot M_{\rm f} c^2/
\rho_{\rm jet} A)^{1/3}$, where $\dot M_{\rm f}$ is the accretion rate of the feedback scheme, $A \sim 7.3$ kpc$^2$ is the area through which the jet is injected, and $\rho_{\rm jet}\sim2 \times 10^{-26}$ g cm$^{-3}$ ($\sim 10^{-1}$ the initial central gas density). We fix the temperature
of the jet to low values ($10^7$ K) in order to keep injected thermal energy on negligible levels compared to
the kinetic flux.

\section[]{Results}

We describe in this Section the results for various flows, exploring a large
set of feedback parameters.
In order to understand the long term behaviour
of the models we have evolved them for 7 Gyr.
The numerical resolution adopted does not allow a deep study of the disturbances
generated by the outflows (such as cavities or shocks). On the contrary, we believe these
models are appropriate to investigate the global properties of the flows, such as the
cooling rate and the azimuthally averaged density and temperature profiles. 
The suite of simulations described here is used to thoroughly explore
the outflows parameter space
in order to bound the region of successful models.

\subsubsection[]{Pure cooling flow}

As a reference flow we first calculated a pure cooling flow (CF),
where no AGN feedback was used, shown in Fig. 1.

\begin{figure}
\centering
\includegraphics[width=55mm]{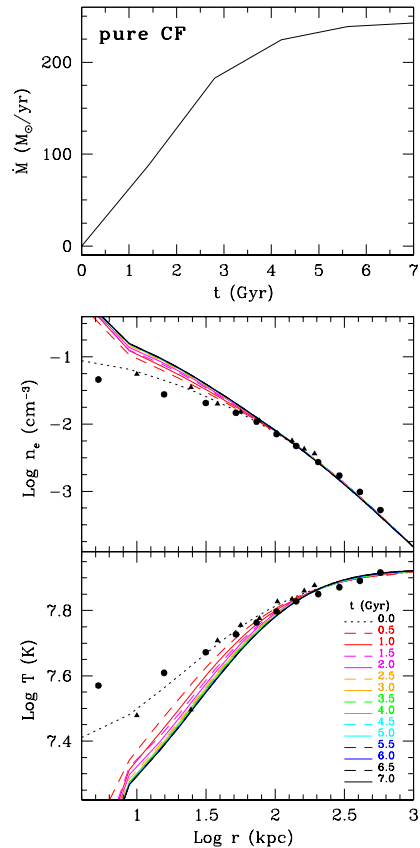}
\caption{Evolution of model CF (no AGN feedback). In the top panel is shown
the gas cooling rate versus time.  The middle and bottom panels show
the temporal evolution of the gas (electron) number density and mass weighted temperature profiles,
respectively. The profiles are displayed at 15 different times, as indicated
in the lowermost panel. Observational data of A 1795 are shown with
filled circles (\citealt{tam01}, {\it XMM-Newton}) and filled
triangles (\citealt{ett02}, {\it Chandra}).} \label{fig:cf} 
\end{figure}

As expected, both density
and (mass weighted) temperature profiles
steepen in the central region, an effect caused by radiative
losses and the consequent subsonic gas inflow. In $\sim 1$ Gyr the
calculated profiles disagree
with the observations of Abell 1795, although not by a great extent.
It is interesting, however, that the logarithmic slope in the core region
of the model temperature profiles at late times is
very close to 0.4, which Sanderson et al. (2006)
found to be typical in the central region of cool core clusters.
The similarity between the temperature distribution in our
pure cooling flow run and real clusters, where heating is
currently preventing gas cooling, put severe constraints
on the feedback process: it must not greatly perturb the
temperature profile shaped by radiative cooling. This is
a demanding requirement (see also \citealt{brm02,brm03}). 

After few Gyr the flow reaches
an approximate steady state (see also \citealt{etb08}
for a quantitative description
of the temporal change in the observable profiles). The bolometric 
X-ray luminosity
slowly increases with time, from $L_x \sim 1.5 \times 10^{45}$ erg s$^{-1}$
at $t=0$, up to $\sim 2.3 \times 10^{45}$ erg s$^{-1}$ at $t=7$
Gyr. The growth of the gas density in the cluster core is responsible
for the increase of $L_x$.

It is interesting
to investigate the global energetic budget. The internal energy within
$r_{\rm vir}$ decreases by $\sim 5 \times 10^{61}$ erg (that is,
of about $1$ \%), while the
potential energy drops by $\sim 4 \times 10^{62}$ erg, considering both the
hot gas remaining in the grid
and the cooled gas at the centre of the cluster.
The kinetic energy
is $\sim 2 \times 10^{58}$ erg and is therefore negligible.
Thus, energy is radiated away ($E_{\rm rad}\sim 4.5 \times 10^{62}$ erg in 7 Gyr)
mainly at the expense of the potential energy of the ICM.

The gas cools in the very center and is removed from the computational
grid by the dropout term described in Section 2.
The cooling rate increases with time, reaching $\sim 250$
M$_\odot$ yr$^{-1}$ at the end of the calculation, a blatant
discrepancy with the observational results (\citealt{tam01,ett02}).
This is of course the
so-called cooling flow problem, the focus of this work.
All the gas in this simulation cools in the very centre of the cluster.
This findings are in excellent agreement with BM06, showing that these
results do not depend on the hydro-code, the coordinate system, the symmetry
or the numerical resolution adopted.

In the following we describe the models with jet feedback. 

\subsection[]{Feedback A ($\bmath{\Delta M_{\rm cool}}$ regulated)}

\subsubsection[]{Model A1, $\epsilon = 5 \times 10^{-4}$}

We start illustrating results for 
feedback A models, where the mechanical
energy of the jet is linked to the gas mass cooled (within $r=10$ kpc).
In Fig. 2 we show the relevant properties for model A1, with
efficiency $\epsilon = 5 \times 10^{-4}$. The width and length of the active region
are 2.7 and 6.8 kpc (1 and 2.5 grid points), respectively.

\begin{figure*}
\includegraphics[width=105mm]{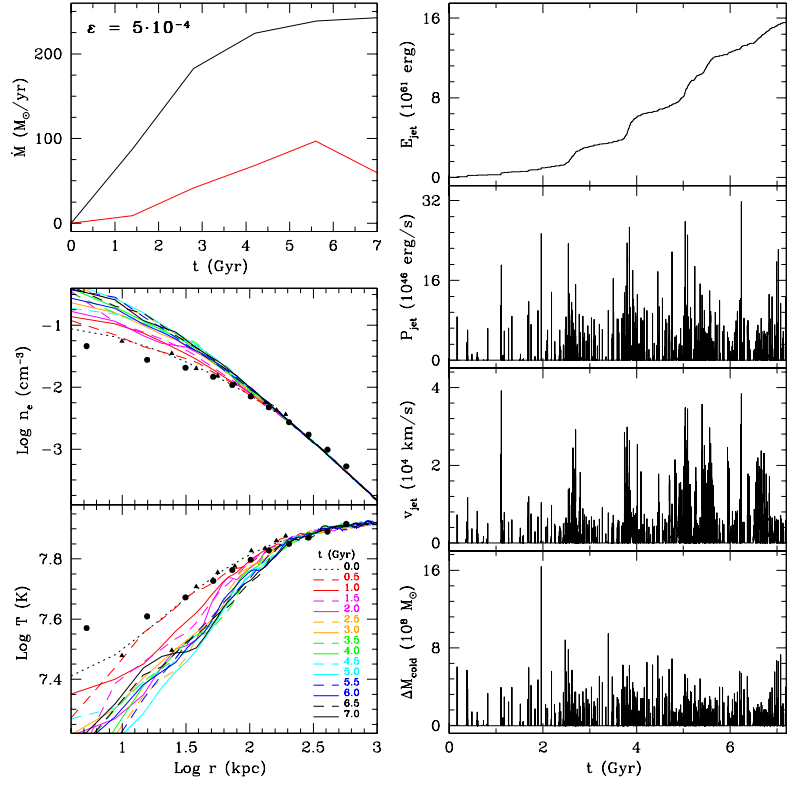}
\caption{Evolution of model A1 ($\epsilon = 5 \times 10^{-4}$). The left column plots
are analogous to those in Fig. 1. In the right column are shown the temporal
evolution of the total kinetic energy injected by the outflows;
the instantaneous power of the outflows; the instantaneous velocity of the outflows;
and the mass cooled in a single timestep. These quantities are calculated for the
half-space $z\ge 0$.} \label{fig:a1}
\end{figure*}

\begin{figure*}
\includegraphics[width=105mm]{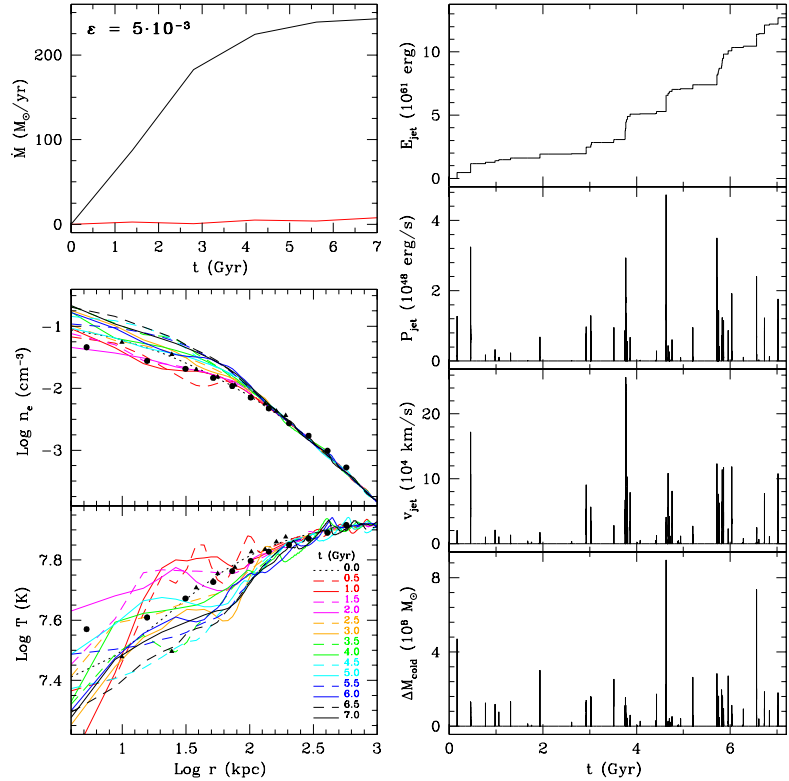}
\caption{Evolution of model A3 ($\epsilon = 5 \times 10^{-3}$). Plots description analogous to Fig. 2.} \label{fig:a3} 
\end{figure*}

The cooling rate, 
although reduced with respect to the CF model, is still
too high ($\dot M_{\rm cool} \sim 75$ M$_\odot$ yr$^{-1}$ at the end of the simulation). 
The azimuthally averaged, mass weighted temperature and density profiles
are similar to those of model CF, with only a little temporal variation:
the weak outflows perturb only slightly the temperature profiles.

In the right column of Fig. 2 are shown the physical characteristics
of all jet events. In these plots the quantities refer to the
half-space $z\ge 0$ considered in our simulations.
In this model the outflows are activated
frequently, because their relatively low mechanical power cannot
prevent the cooling for a long time.

In the upper panel of the right column is plotted the cumulative
(mechanical) energy injected by the jets. At $t=7$ Gyr, $E_{\rm jet}\sim 1.5\times 10^{62}$
erg has been injected in the ICM (in the $z\ge 0$ space). Jets become more frequent
at late times, because of the slow secular decrease of the central cooling time, 
a result of the slight predominance of radiative losses over heating.
At the end of the simulation
$\sim 1.6 \times 10^{11}$ M$_\odot$ have cooled and dropped out of the hot phase.
Clearly, if all the cooled gas were accreted on the central black hole,
as we have assumed in our simple feedback scheme, the final black hole mass would
result far in excess to that of real black holes.
Of course, we could formally avoid the problem of the excessive black hole mass
by assuming that only a fraction of the cooled gas actually accretes on it with
a higher heating efficiency. For instance, this model would be
identical if we assume that only 1\% of the cooled gas is accreted by the central
black hole and the efficiency is increased to $5 \times 10^{-2}$.
This degeneracy means that our model is too simple to allow a proper investigation
of the black hole growth.

In the remaining panels are displayed the power, velocity and mass
cooled (within 10 kpc) during every jet event. Typically, the gas in
an outflow is ejected with $v_{\rm jet} \approx 10^4$ km s$^{-1}$ and power
$P_{\rm jet} = 0.5 M_{\rm act} v_{\rm jet}^2 /\Delta t \approx 10^{46} - 10^{47}$ 
erg s$^{-1}$. $\Delta M_{\rm cold}$, shown in the bottom 
right panel of Fig. 2, is the mass cooled within 10 kpc during a given
timestep, has typical values $10^8 - {\rm few}\, 10^8$ M$_\odot$. 
We note that the Eddington luminosity,
$L_{\rm Edd}\sim 1.5 \times 10^{38}(M_{\rm BH}/M_\odot) \sim 1.5 \times 10^{47}$
erg s$^{-1}$ for a $10^9$ M$_\odot$ black hole, is close to the 
mechanical power of a typical outflow.

In summary, model A1 ($\epsilon = 5 \times 10^{-4}$) resembles a 
pure cooling flow model, with
a cooling rate still too large and must therefore be rejected.
The next logical step is to increase the efficiency $\epsilon$
in order to reduce the mass of the cooled gas and check if more
powerful outflows perturb the variable profiles in an acceptable way.

\subsubsection[]{Model A2, $\epsilon = 10^{-3}$}

Model A2, with $\epsilon = 10^{-3}$, is not shown here and we limit to
a brief description of the results. It has
good temperature profiles, peaked density profile and 
a cooling rate of $\sim 30$ M$_\odot$ yr$^{-1}$ at $t=7$ Gyr, 
which is only marginally acceptable. The cooling rate for $t \lta 2$ Gyr,
however, is $<10$ M$_\odot$ yr$^{-1}$. Evidently, a low efficiency feedback
is able to suffocate the cooling flow for several Gyr. Only at late times
($t \gta 6$ Gyr) the cooling rate becomes too high.
This result emphasizes
the importance of calculating a model for many Gyr to check
the long term thermal evolution. 
With respect to model A1, the jet events are more separated in time, 
especially at early times, consistently with the low cooling rate
at that epoch. The total amount of the energy transferred to the ICM is 
similar to that for model A1, an indication of the self-regulation
of the feedback process.

\subsubsection[]{Model A3, $\epsilon = 5 \times 10^{-3}$}

The increase of the efficiency to $\epsilon = 5 \times 10^{-3}$ generates
a quite successful model (A3), which we discuss in more length (see
also Section 4 for an analysis of the flow dynamics).
The cooling rate (Fig. \ref{fig:a3}) is very low at any time
($\dot M_{\rm cool} \sim 5$ M$_\odot$ yr$^{-1}$), a value fully compatible
with the current observations  ($\dot M_{\rm cool} \la 30$
M$_\odot$ yr$^{-1}$; \citealt{pet03,bre06}).

The azimuthally averaged density and temperature profiles for $t\ge 1-2$ Gyr
are always in good agreement with those observed in A 1795.
Both the ICM density and
temperature vary somewhat with time, following the outflow cycles as expected,
but the cluster always keeps the status of `cool core cluster',
even though gas is essentially
not cooling. The shock waves generated by the outflows (see Section 4)
are not strong
enough to significantly heat the gas and perturb the global, azimuthally
averaged temperature 
positive gradient, although
small amplitude ripples are seen in the profile,
corresponding to weak shocks associated to the jet propagation.
These waves are visible up to a distance of $\sim 400$ kpc.

In this model the AGN feedback is triggered less frequently,
and only about 50 jet events occur. The duty cycle is $\sim 6\%$ (of total time).
The total energy injected,
$\sim 1.25 \times 10^{62}$ erg
is again of the same order as in the previous simulations, and the
average power and velocity of the single outflow is therefore larger. This
energy must be compared with the total energy radiated away, $\sim 1.1
\times 10^{62}$ erg (again calculated in the half-space $z\ge 0$). 
Not surprisingly the two energies are of the same order. In fact,
if the gas density distribution is similar to that of the standard cooling flow model,
then the energy radiated away is also similar. So the energetic balance to stop
the cooling requires that AGN provides an energy $\sim 10^{62}$ erg.
The power of the outflows often exceeds the Eddington luminosity.
While the latter is not strictly relevant in this context, the unpalatable large
power might indicate that this feedback fails to simulate the real accretion process
at work.

It is instructive to examine the evolution of the energetics in the cluster core region
($r \le 50 $ kpc), the most perturbed by the AGN feedback.
After every AGN outburst the core kinetic energy increases on average by 
$\approx 8 \times 10^{60}$ erg (with the most powerful outflows depositing
$\gta 2 \times 10^{61}$ erg), which is comparable to the thermal
energy content of the core region ($E_{\rm th, core} \sim 1.5 \times 10^{61}$ erg). 
However, the kinetic energy generated by an outflow is dissipated in only
$\sim {\rm few} \times 10^7$ yr. Thus, for most of the time the thermal energy
dominates over the kinetic energy.
Following the dissipation of the kinetic energy, the thermal energy also rises,
because of the shock heating. Finally, a large fraction of the thermal energy 
gained by the AGN outburst is transformed in potential energy, as a consequence of
the quasi-adiabatic cooling due to the expansion of the ICM.

In order to understand the effect of the gravity of the central
galaxy, ignored in the models described above, we have rerun model A3
including its contribution. As expected, being the galactic gravity dominant only
in the inner $\sim 20$ kpc, the basic results (cooling rate
and variable profiles) are similar to run A3, and we do not show plots
for this model. The only remarkable difference with model A3 is a reduced frequency
of the jet, compensated with a larger average power. At the end of the simulation
the total energy injected by the outflows in these two models is almost identical.

\subsubsection[]{Model A4, $\epsilon = 10^{-2}$}

When the efficiency is increased further ($\epsilon = 10^{-2}$, model
A4, Fig. 4) the overall results are very similar to those for run
A3.
Again the jet heating generates fluctuations in the temperature
which exceed the ones observed now in A 1795. The time averaged
profiles (not shown), however, agree very well to the observations. 
At almost any time  this model would be classified as a cool core
cluster.

\begin{figure*}
\includegraphics[width=105mm]{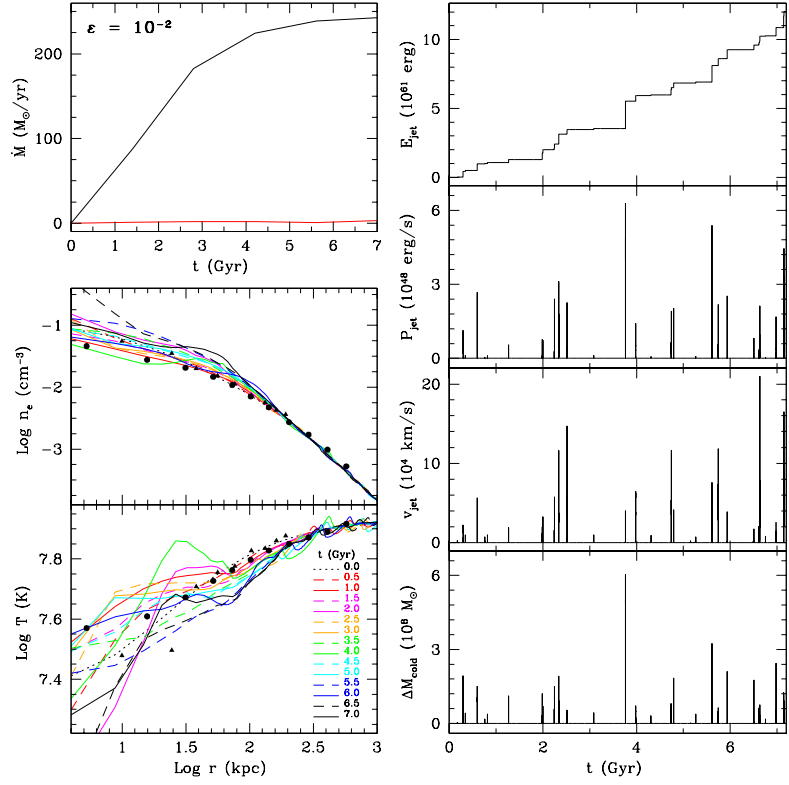}
\caption{Evolution of model A4 ($\epsilon = 10^{-2}$). Plots description analogous to Fig. 2.} \label{fig:a4} 
\end{figure*}

The number of jet
activations during the 7 Gyr of evolution is further reduced, with
only a few happening within the first several Gyr. The total energy
generated by the feedback process is $\sim 1.5 \times 10^{62}$ erg, again 
an acceptable value considering that the total `available' BH energy
is around $1.8 \times 10^{62}$ erg ($E_{\rm BH}\simeq 0.1\,M_{\rm BH}\,c^2$, with
$M_{\rm BH}\sim10^9$).

\subsubsection[]{Model A5, $\epsilon = 5 \times 10^{-2}$}

This model (A5, $\epsilon = 5 \times 10^{-2}$, Fig. 5) clearly
demonstrate the flaws of a too powerful feedback, albeit the total
energy generated by the AGN is once more approximately the same
as in all the other models. Now the 11 AGN outbursts occurring during
the 7 Gyr evolution are quite violent, with power typically larger
than $10^{49}$ erg s$^{-1}$ (right column of Fig. 5). Notice as the outflow
velocities approach the relativistic regime. As expected, the gas
cooling is essentially zero, but the shock heating is too strong and the
central temperature is too high for most of the time. With such an
efficient feedback cool core clusters would be a rarity, in contrast
to the observational evidence.

\begin{figure*}
\includegraphics[width=105mm]{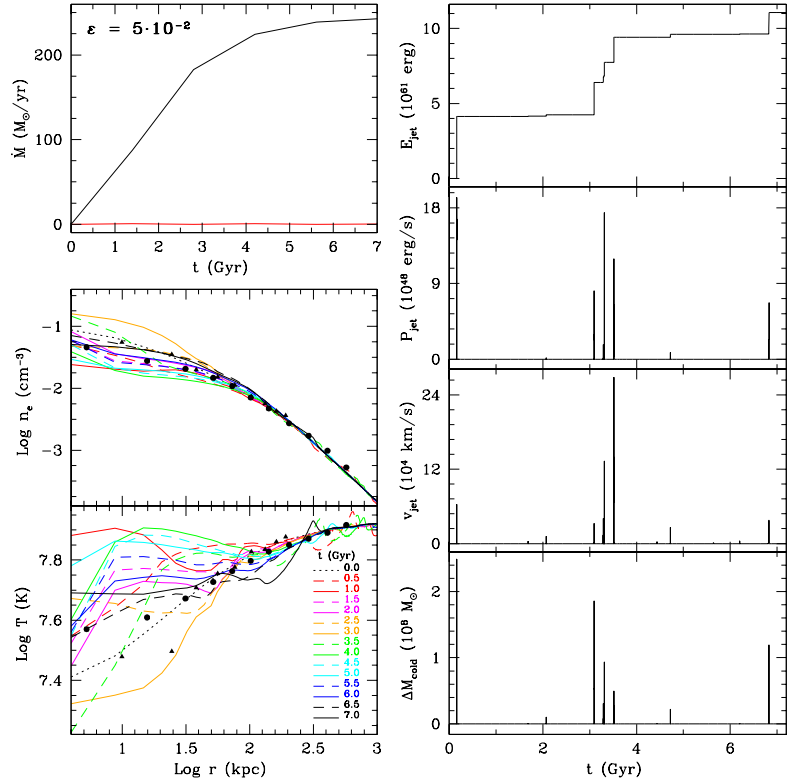}
\caption{Evolution of model A5 ($\epsilon = 5 \times 10^{-2}$). Plots description analogous to Fig. 2.} \label{fig:a5} 
\end{figure*}

\subsubsection[]{Model A6, $\epsilon = 10^{-1}$}

A feedback with  $\epsilon = 10^{-1}$ (an implausibly large
efficiency, but such a model has been calculated for pedagogical 
reasons; model A6, 
not shown here) soon erases the initial cool core
and the system would show a flat temperature profile thereafter
(apart an intermittent mini cool core, $\sim 10$ kpc in size,
associated with the central galaxy and expected also in non cool core
clusters; \citealt{brm02,sue07,sun09}).
In passing, we note that the ubiquitous presence of these galactic scale cool cores
poses very strict constraints on the nature of the AGN feedback, which can
not deposit a large amount of energy in the region near the central
black hole.

\subsubsection[]{Summary of Feedback A ($\Delta M_{\rm cool}$)}

In this section we have illustrated the global features of
feedback A models, in order to check
whether and when non relativistic outflows are a tenable
mechanism for AGN feedback, able to shut off the gas cooling
and at the same time preserving the `cooling flow' appearance ($T(r)$ and $n(r)$).
To summarize the main results,
we find that the efficiency must be in the range
$10^{-3} - 10^{-2}$ in order to generate successful models.
Outflows must be relatively infrequent and powerful.
The total energy necessary to (almost) halt gas cooling
for 7 Gyr of evolution is about $1.5 \times 10^{62}$ erg, not
surprisingly of the same order to the energy radiated away.
However, the energetic balance {\it feedback energy $\approx$ energy radiated
  away} does not guarantee the success of a particular model (see
models A1 and A5, for instance).
Instead, it is crucial the way this energy is communicated to the ICM,
with the appropriate time-scales and power.

\subsection{Feedback B (intermittent) and C (continuous)}

We now illustrate the results when the jets are intermittently activated
at predetermined times (feedback scheme B) or by assuming a continuous outflow
(feedback C).

In run B1, outflows are generated every 200 Myr by setting $v_{\rm jet}=10^9$ cm s$^{-1}$
inside the active region, for a time of 20 Myr each. The results are shown in Fig. 6, rightmost column.
The general features of this model are similar to those of run A2. The cooling rate
is very low for the first $\sim 3$ Gyr and then increases slowly up to $\sim 80$ $\msun$
yr$^{-1}$ at $t=7$ Gyr. The variable profiles remind those of a pure cooling flow calculation,
especially at late times. This model is acceptable for the  first few Gyr.
However, the central gas density
slowly grows with time until radiative cooling prevails over heating, causing the
cooling rate to surpass the threshold of acceptability.
The total energy injected by the outflows is $\gta 2\times 10^{63}$ erg, a much larger value
than in successful simulations adopting feedback scheme A, like A3. This huge amount of energy 
can not come from a single black hole of typical mass $\approx 10^9$ $\msun$.

\begin{figure*}
\includegraphics[width=127mm]{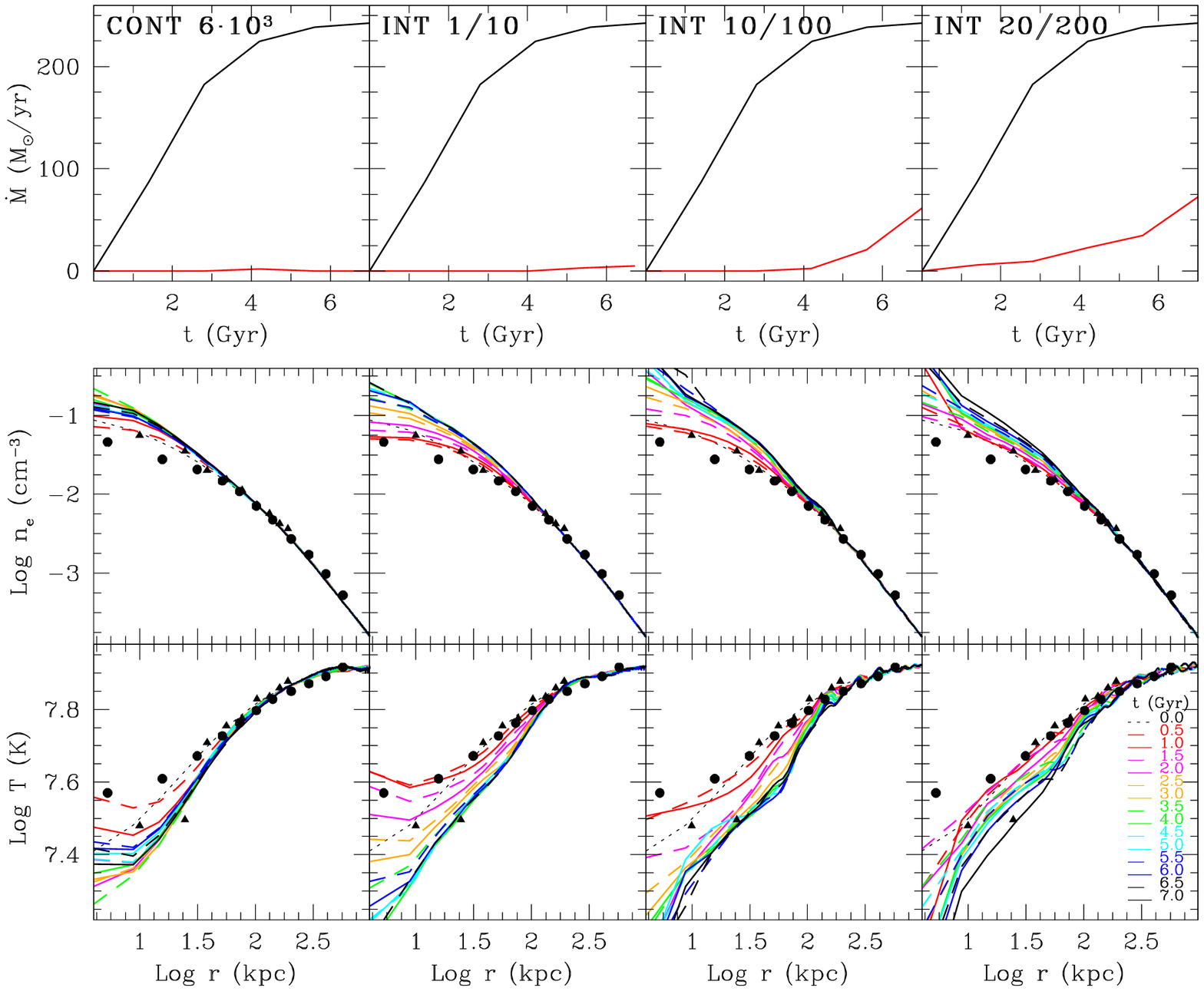}
\caption{Evolution of models C2, B3, B2 and B1, from left to right. Plots description analogous to Fig. 1.} \label{fig:b1b2c2} 
\end{figure*}

Some of the problems of model B1 can be cured increasing the frequency of the jet events
as shown in run B2 (second column from the right in Fig. 6).
This model generates outflows every 100 Myr, each one lasting 10 Myr. Their velocity
is the same as in run B1.
Although the gas cooling rate is much reduced up to $t=5$ Gyr, it rapidly increases
in the last $\sim 2$ Gyr.
The temperature and density profile gradients
become too steep after $\sim 1-2$ Gyr, turning very similar to those of model CF.
While these global properties would make this model acceptable, the required total
energy is again $\sim 2\times 10^{63}$ erg, a value difficult to justify.

A better model can be obtained increasing the frequency of the AGN feedback 
even further. Model B3, second column from the left in Fig. 6, illustrates a run
with outflows activated every $10^7$ yr (similar to the outbursts frequency estimated
for Perseus; \citealt{fab06}), each one persisting for $10^6$ yr. 
As in models B1 and B2, the velocity is set to $10^4$ km s$^{-1}$.
Only a very small amount of gas cools, and the variable profiles resemble again
those of a classical cooling flow. The total feedback energy, $\sim 1.6 \times 10^{63}$ 
erg, is still inconveniently large.

The left column of Fig. 6 shows the outcome of a simulation in which the feedback
is always active (scheme C). Here the outflow velocity is $v_{\rm jet} = 6000$ km s$^{-1}$ (model C2).
This is clearly an extreme model that we have calculated mainly for pedagogical reasons.
The ICM in the centre is continuously heated and transported to large radii,
and almost no gas is able to cool. The central density rises secularly
and it is possible that
some cooling would happen had we evolved the cluster for a longer time.
The temperature profiles agree very well with observations, with a flattening due 
to shock heating visible in the inner $\sim 10$ kpc.
Another unpleasant feature of model C2,
besides the excessive injected energy,
is that the continuous outflow forms a tunnel in the ICM
where it propagates without inflating any cavity (see also the discussion of the
Bondi accretion models in Section 3.4).

We have experimented the effect of varying the outflow velocity in feedback C simulations
(not shown here).
Increasing the jet velocity to $v_{\rm jet} = 10^4$ km s$^{-1}$ (model
C3) leads to negative
temperature gradient in the inner $\sim 10$ kpc, with $T(0)\sim 5 \times 10^7$ K, which
may be in contrast with the observations of \citet{sun09}.
Conversely, with a reduced jet velocity $v_{\rm jet} = 2\times 10^3$ km
s$^{-1}$ (model C1, also not shown here)
the gas cooling
can not be halted and this cluster resembles a pure cooling flow.

In summary, it is possible to find partially successful models using feedback B or (especially) C,
but it is difficult to justify the large amount of outflows energy required or the extreme
character of a continuous jet, when we clearly see the signature of intermittent jets
and AGN feedback
in the different generations of X-ray cavities in many clusters
(\citealt{mcn07,wis07,fab00}).

\subsection[]{Role of the jet size}

Finally, we address how the results depend on the size of the jet active region. 
We increased its length
to $\sim 17$ kpc (the width being the same as before, 2.7 kpc) and calculated several models
varying the efficiency. We do not show figures for these runs, but only briefly discuss
their essential properties.
Overall, we find that the results are very similar to those described in the previous section. 
When the efficiency is $\epsilon = 10^{-3}$ (A2L model), we find that the cooling rate grows
steadily with time, reaching $\dot M_{\rm cool} \sim 80$ $\msun$ yr$^{-1}$ $t=7$ Gyr.
The temperature and density profiles resemble those for run A2, although they are
slightly smoother for  $r\lta 20$ kpc, as expected given that the outflow shocks are generated
at the tip of the source region, located at $z\sim 17$ kpc.
The run with $\epsilon = 5 \times 10^{-3}$ (A3L) has excellent attributes, with very low 
$\dot M_{\rm cool}\lta 10$ $\msun$ yr$^{-1}$ at any time, and smooth profiles in very good agreement
with those observed for A 1795. Finally, we find that high efficiency $\epsilon = 10^{-2}$ still
generates an excellent model (A4L), with very little cooling and very good density and temperature profiles,
where a small temperature bump at $r\sim 30$ kpc, due to the jet shock, is sometime visible in the
mass weighted, azimuthally averaged profile.

At the other extreme, we run a model (A3S) in which the jets are generated by imposing a mass, momentum
and kinetic energy flux in a small region $\Delta x = \Delta y = 2.7$ kpc
at the centre of the boundary plane $z=0$. Again the outflows are triggered when gas cools, with a given efficiency.
The velocity of the inflow is set by the requirement that the kinetic energy injected in a given
timestep is $\epsilon\Delta M_{\rm cool} c^2$ (see Section 2.2). This scheme,
similar to that adopted by \citet{ver06} and \citet{hei06}, 
does not directly change the ICM velocity, but the hot gas is pushed from below by the outflow, which
enters the grid at $z=0$. With efficiency $\epsilon = 5 \times 10^{-3}$, the same as model A3,
we find that the cooling rate (not shown)
is acceptable ($M_{\rm cool}\lta 10$ $\msun$ yr$^{-1}$)
until $t\sim 5.5$ Gyr, then increases up to $\dot{M}_{\rm cool}\sim 60$ $\msun$ yr$^{-1}$ at the end
of the calculation. The temperature profiles often show a positive central gradient in agreement
with the observations of A 1795, although a sharp central 
temperature peak in the inner $\sim 10$ kpc
is present for $\approx 20$\% of the time. We did not pursue the search of the best
model using this outflow generation method, by fine tuning the efficiency or other parameters.

We conclude that the size of source region is not a key parameter of
our models when the feedback power is linked to the cooling gas, as in the runs presented
in Section 3.1.

\subsection[]{Feedback triggered by Bondi accretion}
As discussed previously, the feedback schemes adopted for the models presented in
Section 3 are just a convenient way to link the response of the black
hole to the cooling of the ICM, in order to make the feedback process
self-regulating. The physics of the accretion, black hole growth and
outflow generation, poorly understood in general, is by no means
captured in these simulations. This is a general weakness of current models
of AGN feedback: the energy (or momentum) is injected in the ICM
according to essentially ad hoc prescriptions.

A manifestation of this deficiency is the excessive
accretion rates occurring even in our most successful models using feedback A.
For instance, in model A4 the most powerful jets ($P_{\rm jet}\sim
10^{48}$ erg s$^{-1}$) imply accretion
rates of $\sim 2000$ M$_\odot$ yr$^{-1}$, much larger than the
Eddington rate for a black hole mass of $3 \times 10^9$ M$_\odot$,
$\dot M_{\rm Edd} \sim 60$ M$_\odot$ yr$^{-1}$.
These super-Eddington values are not very common for AGN (e.g. \citealt{kin09})
and yet, according to the
results based on feedback A (Section 3.1), we {\it need} fast and energetic
outflows, which imply accretion of large gas masses,
 to shut down the gas cooling.
 
At this point we tried a different approach to mitigate this too explosive behaviour.
The Bondi accretion theory (\citealt{bon52}), although highly idealized
with respect to the complexity expected in real accretion systems (thin or thick discs, ADAF, etc.),
has been widely used to estimate accretion rates on supermassive
black holes. However, in presence of a standard cooling flow
the `unpertubed' gas is not at rest as in Bondi theory.
The accretion rate on the central black hole is then determined by the
ICM inflow rate and should be proportional to the cooling
rate, which in turn is determined by the global physical conditions
of the ISM or ICM.
Thus, the Bondi rate should not be particularly relevant if the gas cooling rate
is large.

Despite this theoretical consideration, our resolution is still very far from capturing the
real central accretion engine (few tens of Schwarzschild radius) and we have to base
the triggering mechanism to mean large scale values. In this sense it is interesting to calculate a model in
which the AGN feedback is linked to the accretion rate estimated
by the Bondi's formula: $\dot M_{\rm B} = 4\pi (GM_{\rm  BH})^2\rho_0/c^3_{{\rm s}0}$ (see Section 2.2). 
Note that the accretion increases when temperature drops and density grows, or better when the entropy ($s\propto 
T/\rho^{2/3}$) is falling off. Therefore, also Bondi models are sensitive to gas cooling, but in a more gentle and moderate manner.
If such a small accretion rate is sufficient to
halt the cooling, then this model is self-consistent.

In a sample of nine X-ray bright elliptical galaxies \citet{all06}
estimated typical Bondi accretion rates $\dot M_{\rm B}\approx 0.01$ M$_\odot$ yr$^{-1}$. 
For galaxies at the centre of rich clusters the Bondi rates are not expected to
greatly exceed these values. Therefore, the Bondi AGN feedback model
should operate in an opposite regime than that in models A3 or A4.
In particular, we expect a quasi-continuous AGN activity of moderate intensity.
Such a model is similar to that calculated by \citet{cat07}, who were able to drastically reduce the
cooling rate (injecting also much thermal energy), although the price to pay for this result was the absence
of a cool core.

In the first BONDI model the outflow is generated in the same way
as for feedback scheme A, that is injecting the energy to the gas
contained in the active region ($2.7\times 2.7 \times 6.8$ kpc$^3$), but with an efficiency of 0.1. 
Note that here we averaged $\rho_0$ and $c_{{\rm s}0}$
in a radius of 10 kpc from the centre.
We do not show figures for this run, but limit ourself to a general
description.
As anticipated, this model is characterized by an almost continuous
activity of relatively weak outflows, with typical velocity of $\sim 1000$ km
s$^{-1}$. The density and temperature profiles are very
similar to those of the classical cooling flow simulation and the
cooling rate rises to unacceptable values of $\sim 250$
M$_\odot$ yr$^{-1}$. This simulation recalls A1: the very frequent, low power outflows
are not able to prevent massive cooling,
resulting in a model almost
undistinguishable from a pure cooling flow. It is important to note, that 
even if the heating is always turned on, the jet power is greatly oscillating from 
$10^{43}$ to few $10^{45}$ erg s$^{-1}$. In fact the Bondi rate feels the low entropy gas which is
cyclically building in a small torus, around the perpendicular outflow ($\sim 10$ kpc).
In the end this simulation
fails to meet our requisites for a successful model.

In order to partly detach the Bondi accretion rate from the huge amount of cooled gas in the inner tens of kpc (and thus Feedback type A), we need to double the resolution and then halve the averaging zone (5 kpc). 
To explore Bondi feedback further, we wanted also to use the smallest jet possible, calculating
a second model with the alternative jet generation scheme: 
now the jet (mass, momentum, and mechanical energy flux
given by $0.1 \dot M_{\rm B}c^2$) is injected
from the grid boundary at $z=0$ as described in Section 2.2.
 
\begin{figure*}
\includegraphics[width=105mm]{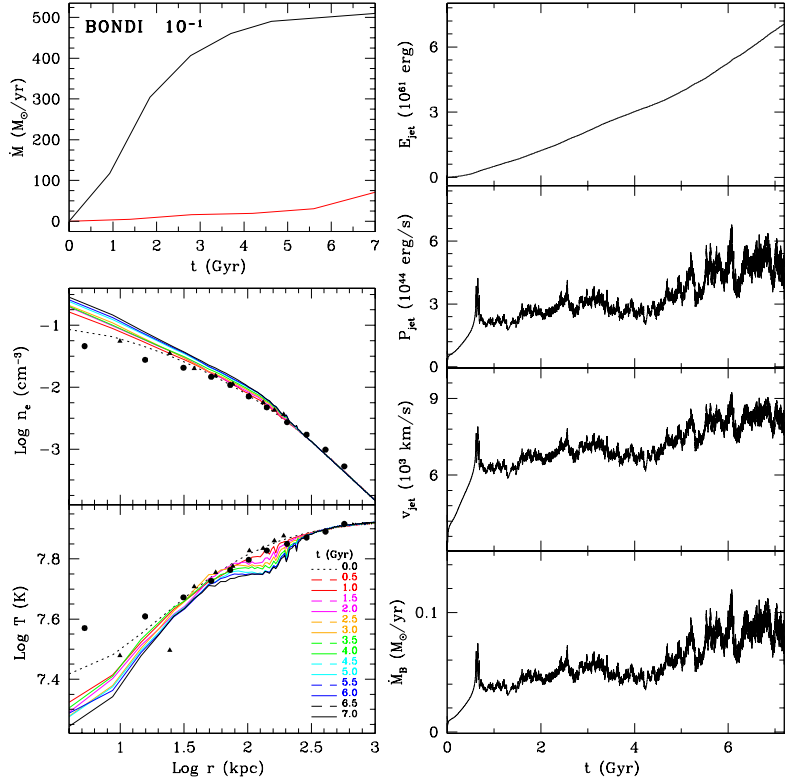}
\caption{Evolution of model BONDI2. Here the jet is injected from the grid boundary at $z=0$ 
(see Section 2.2 for details). In the left column are represented the same quantities as in Fig. 2.
In the right column are shown, from top to bottom, the total energy injected by the AGN,
the instantaneous power of the jet, the velocity of the jet and the Bondi accretion rate,
respectively.} \label{fig:bondi} 
\end{figure*}

The results for this model (BONDI2) are shown in Fig. 7. The $T$ and $n$ profiles and the
cooling rate are good, with the
latter exceeding the observational limits only after $\sim 6$ Gyr. As in the previous
Bondi model the activity of the AGN is continuous and moderately weak, but now the jet power is very steady (on the contrary of previous simulation)
in the range $2-3 \times 10^{44}$ erg s$^{-1}$, 
for a total injected energy of
$\sim 7 \times 10^{61}$ erg. In fact, the accretion rate does not vary much, $\dot M_{\rm B}
\sim 4-8 \times 10^{-2}$ M$_\odot$ yr$^{-1}$, slightly larger than the estimates by 
\citet{all06}. 

In shrinking the central averaging zone (for the entropy) the Bondi accretion does not present many
fluctuations and the triggered outflows
seem capable of reducing the cooling flow. 
This might be more evident with higher resolution (up to Bondi radius) and thinner jets.

Furthermore, comparing both Bondi models, we have noticed that, 
from a numerical point of view, injection through the boundaries seems a better method to produce a jet and subsequent entrainment (as pointed out by other authors, \citealt{omm04} for example). 
This way the jet flux is tightly coupled
to the (PPM) hydrodynamical algorithm and not inserted like a split source term, modifying by-hand
flow variables in cells of the effective computational domain. Moreover, keeping a fixed
jet density, the velocity can not reach very high values (like $10^5$ km s$^{-1}$), 
facilitating the stability of the code and moderate CFL number ($\sim0.4$).

A possible riddle for BONDI2 model is the absence of frequent jet-inflated
spherical cavities (see also model C in Section 3.2).
The continuous AGN activity (mean $v\sim 7000$ km s$^{-1}$) carves a narrow tunnel of about 50 kpc in length,
although its density contrast with the environment is large only for $z\lta 20$ kpc (and not particularly evident in the SB maps). See Section 6.3 for a deep discussion.

We will certainly extend the study of Bondi-type feedback in a dedicated future paper.

\section{Dynamics of Model A3}

In this Section we will discuss the dynamical evolution
of the outflows for one of the best models, A3 ($\epsilon = 5\times 10^{-3}$). Contrary
to the almost steady, and not variegate, evolution of model BONDI2, the density and
temperature maps (i.e. the $x-z$ midplane) of A3 show significant temporal variations. 
Note that in this Section we are analysing {\it physical} quantities, while the right comparison
with observations should be done through {\it emission-weighted} ones. We will broaden this aspect in Section 5.

The first outflows is triggered at $t\sim 160$ Myr (see Fig. \ref{fig:a3}),
with a velocity $v_{\rm jet}\sim 2\times 10^4$ km s$^{-1}$, mechanical power $P_{\rm jet}\sim
1.3 \times 10^{48}$ erg s$^{-1}$ and a total energy injected of $\sim 4.5 \times 10^{60}$ erg.
The large jet power is due to our adopted feedback method A, where outflows last only while
gas is cooling to low temperatures. This implies that the AGN is active only for 
few timesteps (typically $\Delta t \approx 10^5 - 10^6$ yr), since the jet promptly stops
the gas cooling in the central region. During a single timestep large
masses of gas can cool (also a result of the coarse resolution used),
therefore the energy $\epsilon \Delta M_{\rm cool} c^2$,
corresponding to the cooled gas mass $\Delta M_{\rm cool}$, results in a large, and probably
exaggerated, instantaneous outflow power.

\begin{figure*}
\includegraphics[width=107mm]{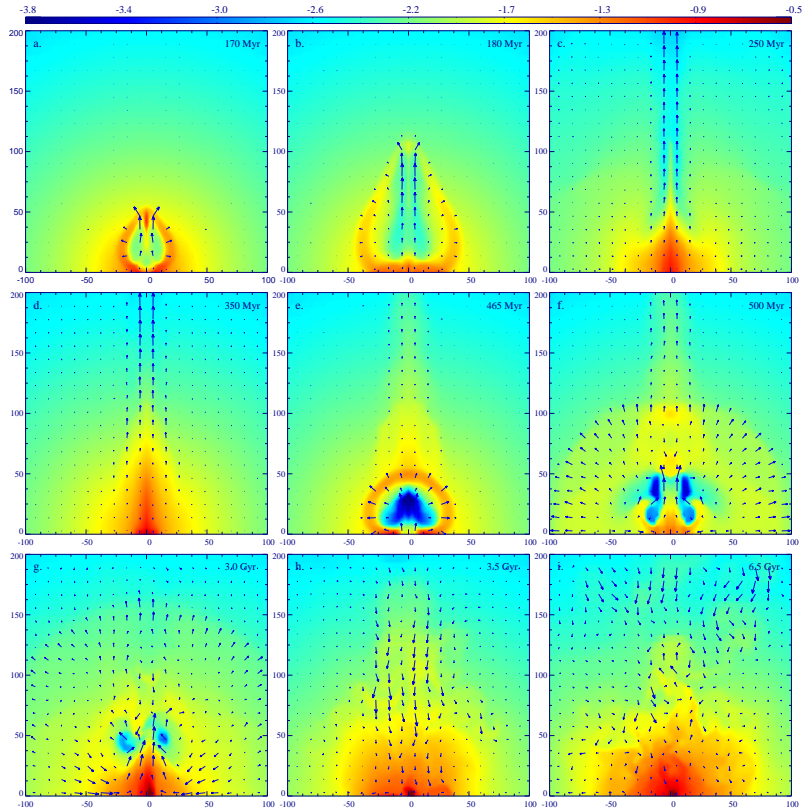}
\caption{A3 model: logarithm maps of electron number density (cm$^{-3}$)
in the $x-z$ midplane (kpc unit), with velocity
field superimposed. The color scale is given by the bar on the top, 
while arrows length normalization varies.
Times are indicated on every top right corner.}
\label{fig:rhomaps} 
\end{figure*}

\begin{figure*}
\includegraphics[width=107mm]{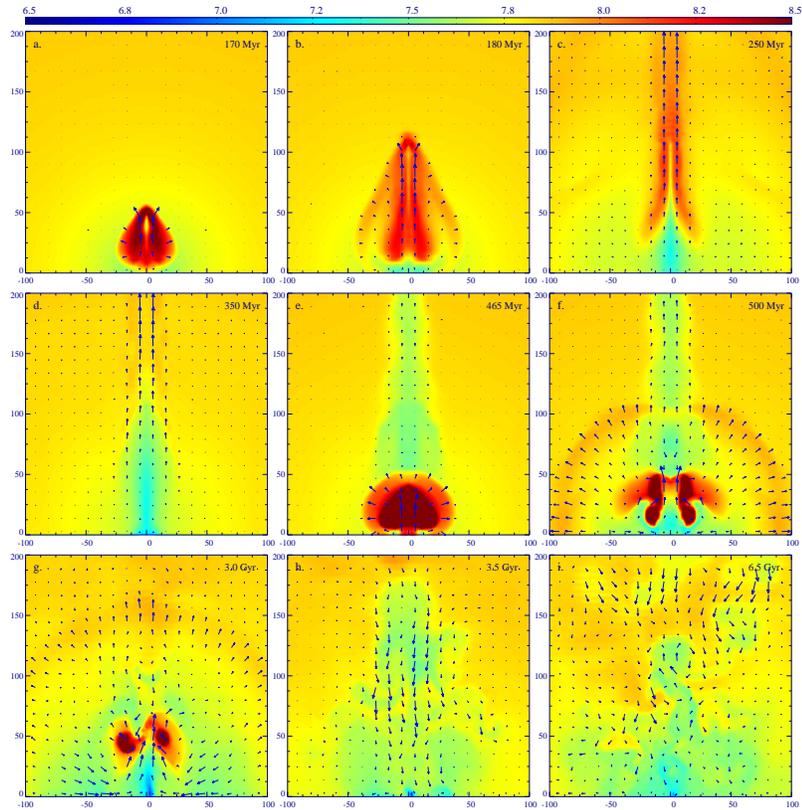}
\caption{A3 model: maps of the logarithm of gas temperature (K). See Fig. 8 for other details.}
\label{fig:temmaps} 
\end{figure*}

In Fig. 8 and 9 are represented density and temperature maps in the $x-z$ plane (the outflows
are generated along the $z-$axis) for the first two AGN outbursts.
At $t=170$ Myr (Fig. 8a), after $\sim 10$ Myr since  the jet started, an ellipsoidal cavity 
with major (minor) semiaxis of about 15 (10) kpc has been carved,
surrounded by a shell of shocked gas, whose density is about twice the ambient density
at nearby locations. The temperature of the shocked gas is $\sim 10^8$ K,
about three times that of the unperturbed ICM.
Therefore the young cavity, expanding approximately
at the sound speed is surrounded by a weak, hot rim.
The cavity has a relatively low\footnote{
We lack the effect of relativistic protons, whose pressure is likely
to be relevant in expanding the cavity (\citealt{mab08a,mab08b,gum10}).}
density contrast with 
the environment, $\sim 3-5$.
Outside the cavity shock the ICM is still slowly flowing in,
as a classical cooling flow. The azimuthally averaged (mass weighted)
temperature profile
(Fig. 10) shows a spike
at $r\sim 30$ kpc, with a fractional increase $\sim 60$\%.
\begin{figure}
\centering
\includegraphics[width=57mm]{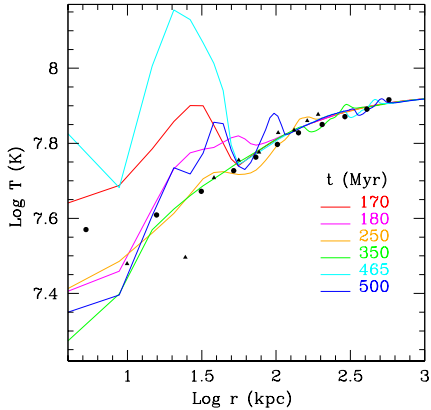}
\caption{A3 model: mass weighted, azimuthally averaged temperature profiles 
for the six times showed in upper two rows of Fig. 9.}
\label{fig:6temp} 
\end{figure}
The cavity expands, increasing its ellipticity, and at $t\sim 180$ Myr (Fig. 8b)
it has approximately a cylindrical shape, extending up to $z\sim
100$ kpc and with a radius $\sim 15$ kpc. The bubble is filled with hot ($\gta 2\times 10^8$ K)
gas raising along the $z$-axis at about its sound speed.
The density contrast is again
$3-4$. A very weak `pear-shaped' shock surrounds the cavity. The density
gradient along the post-shock region, with denser gas closer to the
equatorial $x-y$ plane, makes the low$-z$ part of the shock detectable,
while at large distance from the centre the low density contrast likely
prevents the shock detection (see also the X-ray brightness map in Fig. 11).
The physical temperature jump across the shock for $z\lta 25$ kpc is only
$\sim 25$\% and increases slightly at large $z$. 
The structure qualitatively reminds the one observed in the elliptical galaxy
NGC 4636 (\citealt{fij01,bal09}).
In the azimuthally averaged temperature (Fig. 10) the weak shock and the heated gas are visible
as a small bump $\sim 25$\% in amplitude, located at $10 < r < 70$ kpc.

20 Myr later the cavity lengthens and narrows, while the back flow generates a
dense, relatively cold filament protruding in the cavity (e.g. \citealt{mab08a,gar07}). 
As for the previous times, in most of the volume
of the cluster the ICM is inflowing as in a standard cooling flow, and the 
averaged profiles agree very well with those observed for A 1795;
after 40 Myr since the powerful AGN outburst, the cluster fully restored
the `cool core' appearance it had at the beginning of the calculation.
A very small amplitude `ripple' ($\Delta T/T\sim 10$\%)
is visible in the temperature profile
at $\sim 100$ kpc as the integrated effect of the elongated weak shock.

At $t=250$ Myr (Fig. 8c) the filament reaches $z=100$ kpc, while the 
cylindrical (subsonic) outflows in the (now almost disappeared) cavity
reached $z\sim 250$ kpc. The temperature ripple moves forward at the sound speed
and slowly weakens.

At $t\sim 350$ Myr (Fig. 8d) the cavity disappears while the dense, relatively cold
filament formed by gas formerly at the centre and lifted at large $z$ by the jet motion,
is still clearly visible. The inner part of the filament, at $z\lta 40$ kpc,
reverses its velocity and starts to fall back toward the centre.
This is reminiscent of the kinematic of the emission line filaments
observed in the Perseus cluster (\citealt{hat06}).
The outer region around the $z-$axis is instead still slowly flowing out.
This moment marks a new phase in the cluster lifecycle.
The fall back of relatively dense gas
preludes the next cooling/feedback event, which occurs at 
$t\sim 450$ Myr. The feedback cycle starts again in a qualitatively
similar way.

Fig. 8e shows the density map at $465$ Myr. A new cavity is formed, at first of
approximately spherical shape centred at $z\sim 25$ kpc, radius $\sim 20$ kpc 
and high density contrast ($50-80$).  The almost spherical symmetry of the cavity
is caused by the inflow along the jet axis, whose ram pressure slows down
the expansion along that direction.
The shock, already weak (Mach number $\sim 1.5$) is 
also nearly spherical, with radius $\sim 35$ kpc.
Note that, at this time, the outflow has an energy of $\sim8\times10^{60}$ erg, while
the cavity has roughly $\sim5\times10^{59}$ erg (with the usual $2.5$ $PV$). Thus most of the
mechanical energy is not used to form the cavity.

\begin{figure*}
\includegraphics[width=107mm]{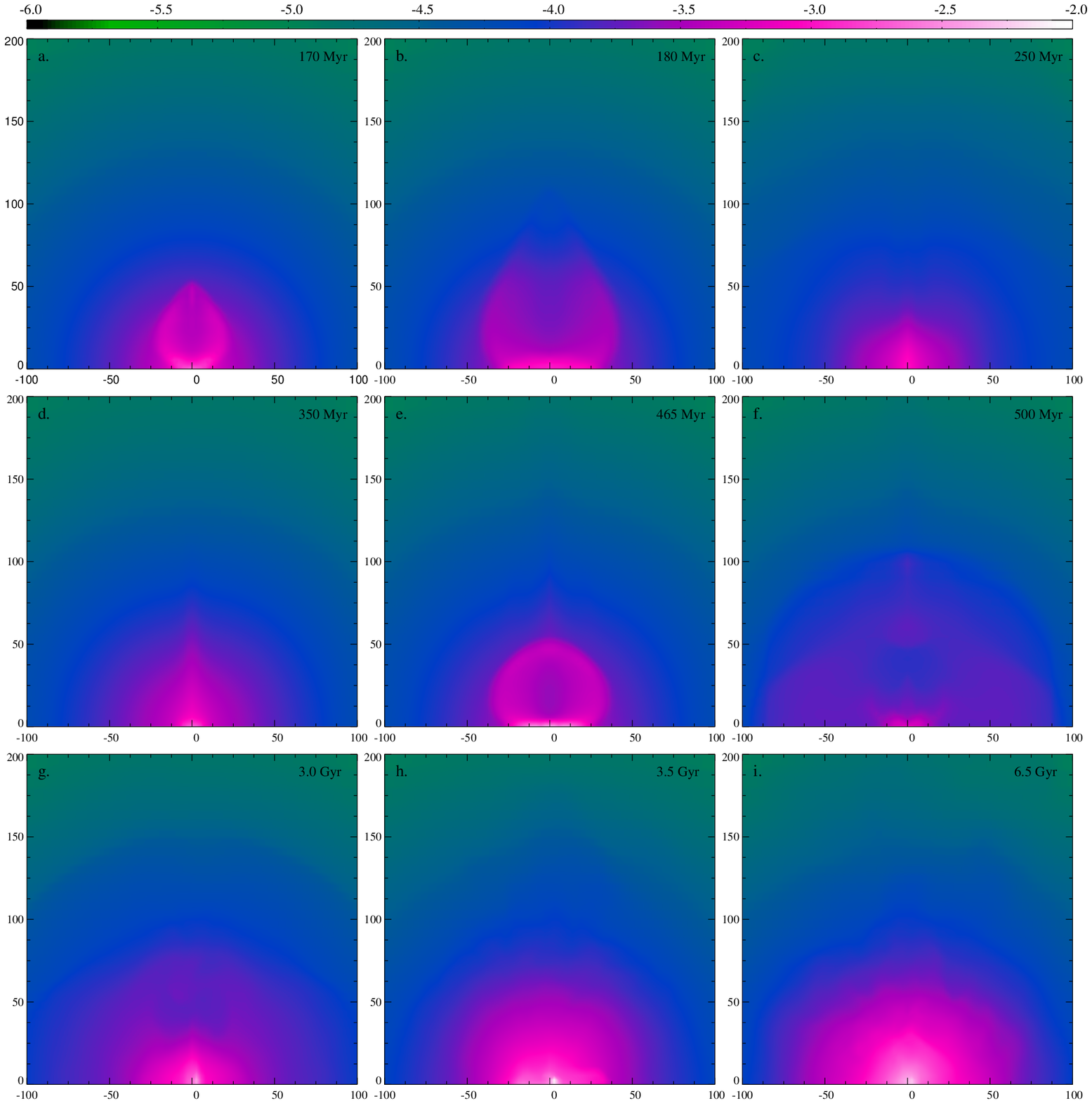}
\caption{X-ray surface brightness maps for model A3 at various times. The $x$-axis is horizontal,
while the $z$-axis is vertical (kpc units).} \label{fig:SBmaps} 
\end{figure*}
\begin{figure*}
\includegraphics[width=107mm]{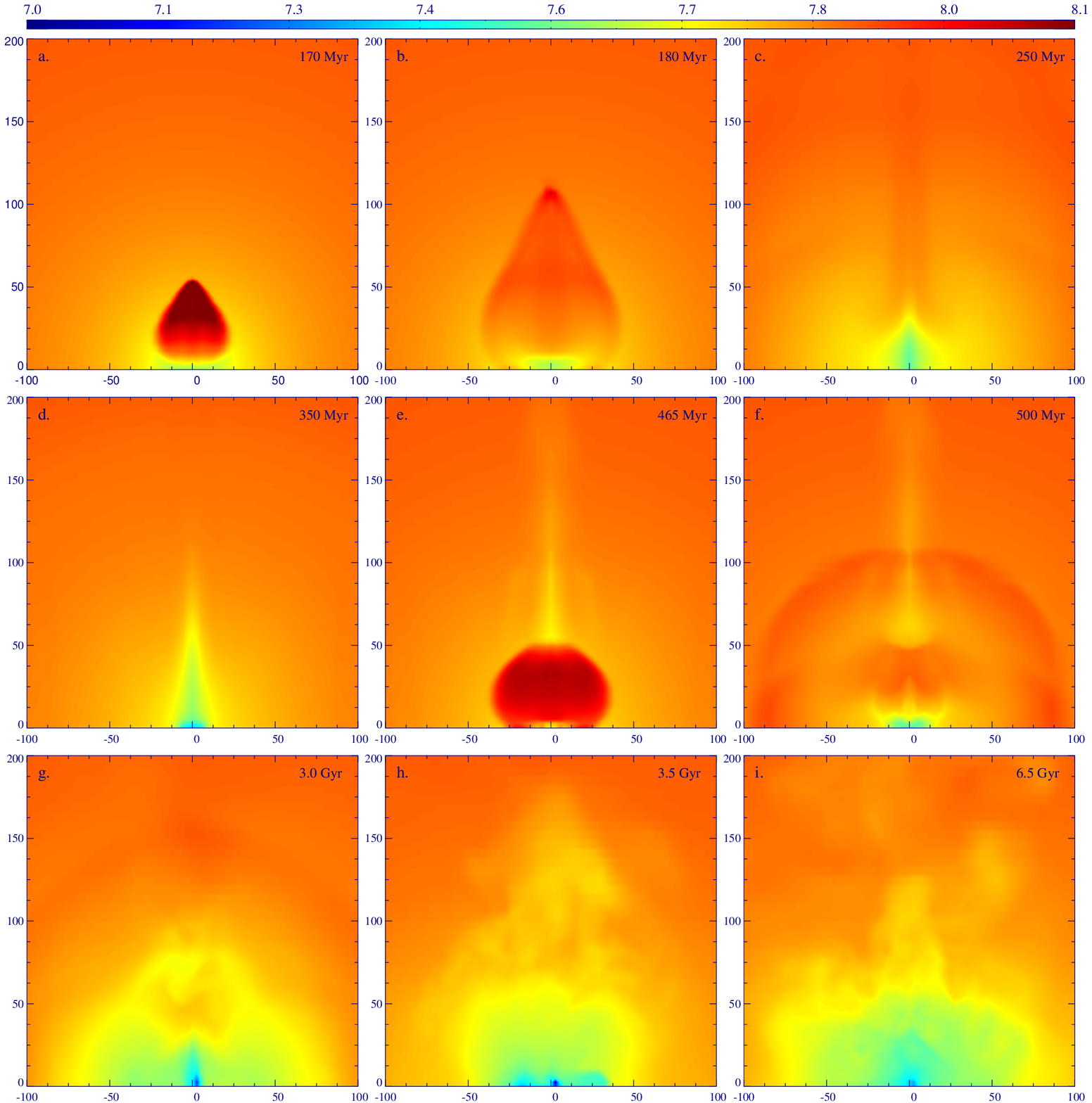}
\caption{Emission weighted temperature maps for model A3 at various times (see also Fig. 11).} \label{fig:Tewmaps} 
\end{figure*}

Again the temperature profile shows the signature of the young AGN outburst
with a strong peak for $r\lta 50$ kpc, approximately the location of the shock
along the $z-$axis. A very weak ripple, vestige of the first jet event, is visible
at $r\sim 400$ kpc.
 
At $t=500$ Myr (Fig. 8f) the cavity shape is strongly affected
by the back flow and the density contrast lowers. The weak shock ($M\sim 1$), slightly
elongated in shape, is now located at a distance of $\sim 100$ kpc, much further
away than the cavity. Within $r\sim 100$ kpc the cluster atmosphere is slowly
moving outward (with velocity in the range $200 - 500$ km s$^{-1}$).
Being the motion very subsonic the density profile changes very slowly.
The outflow is decelerating and in $\approx 100$ Myr reverses its direction
approaching the dynamics of a classical cooling flow.
Thus, the ICM in the cluster core undergoes cycles of slow contraction and expansion,
following the rhythm of the AGN activity.

The cylindrical outflow with average velocity
$v_{\rm z}\sim 300$ km s$^{-1}$
along the $z-$axis, remnant of the previous
AGN outburst is still present, and extends beyond $z=200$ kpc.
At the same time the inflowing gas in the lower part of the filament
is effectively preventing the cavity from expanding or raising buoyantly
above $z\sim 50$ kpc.

Finally, in the bottom row of Fig. 8 we show few snapshots of the density distribution
at late times. As the cycle of feedback proceeds the flow develops a more
turbulent character. In panel 8g is represented the density map at $t=3$ Gyr.
The cavity, generated at $t\sim 2.91$ Gyr, is distorted by the ascending backflow 
and by falling gas in the filament along the $z-$axis. As a result it acquires
the shape of an asymmetrical torus.
As in the previous aftermaths of the jet episodes, the ICM is flowing inward
in most of the cluster volume, and is deviated outward along the $z-$direction when
it reaches $r\sim 20$ kpc.

Panel 8h illustrates the cluster during a quiescent period ($t=3.5$ Gyr),
just before a jet is triggered. The density distribution is very smooth and the cluster
appearance is that of a standard cooling flow (see also the profiles in Fig. 3).
A large, slightly overdense region around the $z-$axis is
falling toward the centre with velocity varying between 50 and 500 km s$^{-1}$.

Finally, in panel 8i we show the density distribution at $t=6.5$ Gyr, again in
an epoch long after an AGN outburst. Again, the density is smooth in the core
region, while shows some variation for $r\gta 50$ kpc. On the
contrary, the velocity field is
chaotic and very subsonic; it promotes mixing of the metals produced
by the SNIa exploding in the central galaxy (see Section 5.3).
Few streams of moderately overdense gas are falling
from large radii $r\gta 100$ kpc with velocity $\sim 300$ km s$^{-1}$.

\section[]{X-ray Observable Predictions}
The $x-z$ cuts through the centre of the cluster are very useful tools to
investigate the intrinsic dynamical evolution of flow variables, like density, temperature
and velocity. However, in astrophysics we are limited to observations based on the surface
brightness (SB) in a typical spectral range (in our case mainly X-ray band). In theory we have to mock
every aspect of the observation, from different atomic emissions to instrument response.
Due to our limited resolution, it is sufficient for us to just integrate the emissivity or emission-weighted quantity
along line of sight ($y$-direction) in an energy range of the X-ray band $\sim 0.5-10$ keV (similar to {\it Chandra}). Note that we have created a parallel code that is able to interpolate every slice of the data cube
(remember that we have 9 different levels) and then perform the above mentioned integration, in order to obtain more precise SB maps. With these maps we can test further the observability of our models and, in the lack of real ones,
just make predictions that could be in the near future verified or falsified. We will also test common
assumptions regarding gas hydrostatic equilibrium, taken in many observational analysis, that seem too restrictive.

\begin{figure}
\centering
\includegraphics[width=57mm]{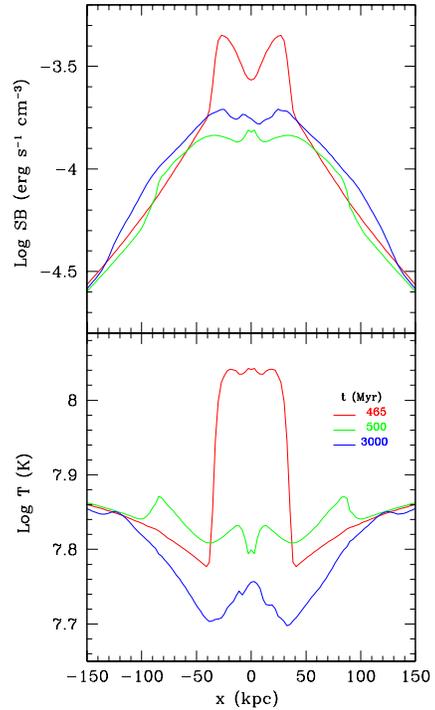}
\caption{A3 model: X-ray surface brightness and emission weighted temperature
1D cuts (of Fig. 11 and 12 maps), through $z=25,\; 50,\; 50$ kpc, 
at 465, 500, 3000 Myr, respectively.} \label{fig:profiles} 
\end{figure}

\subsection[]{Cavities and shocks}
We begin investigating the detectability
of faint features, like X-ray cavities and shock waves, generated
by the AGN outflows. In Fig. 11 we show the X-ray surface brightness
maps for run A3, at the same times showed in Fig. 8 and 9.
The emission weighted temperature map is displayed in Fig. 12.
X-ray depressions are clearly seen at $t=180,\;465,\;500$ and $3000$ Myr.
Evidently, subrelativistic, massive, collimated outflows can generate
cavities with typical diameters of 15-40 kpc. Often the X-ray bubbles are
surrounded by bright rims of shocked gas. Relatively weak waves 
are present at large distance from old cavities, both features being
generated by the same AGN outburst.

To better quantify the brightness perturbations produced by the
outflows we show in Fig. 13 the profiles along the $x-$direction at
$t=465,\; 500$ and $3000$ Myr, taken at $z=25,\; 50,\; 50$ kpc, respectively.
The sharp jumps at $x\sim \pm 30$ kpc at $t=465$ Myr, where  the
cavity shock increases the brightness by a factor $\sim 2.5$ would be
manifestly visible in the X-ray image. The relative central depression
(the cavity) is indeed brighter than the same region before the jet activity.
The emission weighted temperature map reveals that the cavity region
is markedly hotter (by $\sim 80$ \%) than the nearby ICM, 
which is not commonly observed. 
This is a consequence of the infrequent, but explosive outflows like those of model A3.

\begin{figure*}
\includegraphics[width=115mm]{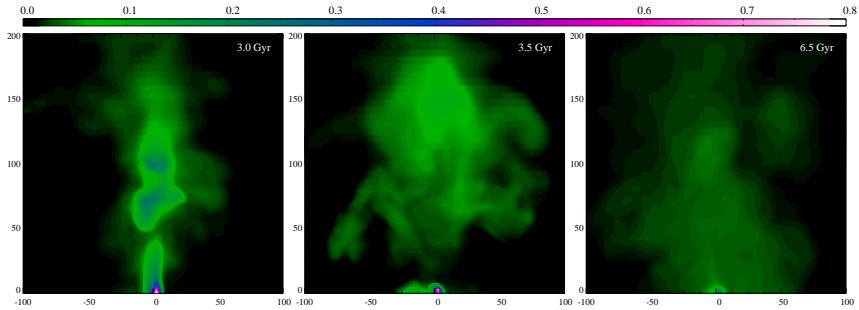}
\caption{Emission-weighted iron abundance maps (Z$_{\rm \odot}$ unit) in the $x-z$ midplane for model A3,
at three different late times.} 
\label{fig:ironmap} 
\end{figure*}

The maps and profiles at $t=500$ Myr show a weak cavity centred
in $y\sim 30$ kpc, with radius $\sim 20$ kpc. This cavity is also
slightly hotter than the surrounding gas. At this time the most
interesting feature is the weak shock located at a radius $\sim 100$ kpc:
Mach number is $\sim1.1$, in good agreement with observations of shocks
driven by AGN (\citealt{bla09}).
The surface brightness profile shows a clear front at the shock
position and the emission weighted temperature jumps of $\sim 10$ \%.

At $t=3$ Gyr another weak cavity, again $\sim 20$ kpc in radius, is
present. The brightness depression is only $\sim 10$ \% and the
temperature is higher than the ambient one by the same percentage.

To summarize, our best models presented in Section 3.1, 
with relatively powerful and infrequent
outflows, have the tendency to produce cavities in a violent
way, generating shocks which heat much the surrounding ICM.
This is also a reason why we investigated the effect of a weaker and
quasi-continuous self-regulated feedback (Section 3.4). We will discuss
the differences in Section 6. 

\subsection{Iron enrichment and mixing}

It is expected that directional outflows would generate metal
inhomogeneities through the ICM. Iron is the most relevant 
and easily measured element, 
being produced mostly by SNIa,
which are still exploding in the giant elliptical at the centre
of cool core clusters.
The same outflows, however, generate turbulence and bulk 
motion, which in turn tend to stir and mix the ICM, restoring
homogeneity and erasing abundance gradients.

We present here a brief analysis of the emission-weighted iron abundance
evolution for model A3. We model only the Fe-enrichment
produced by the SNIa (and stellar winds) occurring in the central galaxy, in the
time interval $6.7-13.7$ Gyr, the latter being the current age of the universe $t_{\rm n}$.
Neglecting the iron produced by the SNII and by the other cluster
galaxies, we are not in the position to investigate the complete
chemical evolution of the cluster (this will be the subject of a
future work). Instead, we are interested here in quantifying
the abundance anisotropies caused by the AGN outbursts.

The current SNIa rate is assumed to be $\sim 0.1$ SNu
(supernovae in 100 yr per $10^{10}L_{B,{\rm \odot}}$), slightly below
to that estimated in local early-type galaxies (\citealt{cap99})
but in agreement with the value necessary to generate
the observed abundance in well observed giant elliptical galaxies
(\citealt{hub06,mab03}).
The time dependence of the SNIa rate is assumed to be 
$\propto (t/t_{\rm n})^{-1.1}$ (\citealt{gre05}).

The iron density is implemented in the code as a tracer of the flow
($\rho_{\rm Fe}=\phi_{\rm Fe}\rho$ with $\phi_{\rm Fe}$ a scalar between 0 and 1),
following the usual advection equation:
\[
\frac{\partial\rho_{\rm Fe}}{\partial t} + \bmath{\nabla}\cdot\left(\rho_{\rm Fe} \bmath{v}\right) = S_{\rm Fe},
\]
where the source term $S_{\rm Fe}$ depends mainly on $\alpha_{\rm SN}\rho_{\ast}$ (Sec. 2), 
as previously discussed (see \citealt{mab03} for more details).

In Fig. 14 we present the $Z_{\rm Fe}$ maps for model A3,
at three different late times. Note that the background (black) 
is zero, but in reality should be around 0.3 Z$_{\rm \odot}$: here 
we are just interested in the contrast.
At 3 Gyr (first panel) the abundance is highly asymmetric. A powerful jet
event has recently occurred (see Fig. 3), therefore the metals created
in the central elliptical galaxy (within few kpc) are transported out along the $z-$direction
up to a distance of $150-200$ kpc. This is an unmistakable mark of the
presence of a powerful AGN outflow. In fact we suggest that iron abundance can be used as
a reliable tracer for any AGN jet-outflow activity, instead of common entropy maps.
At the centre the abundance is about 0.4-0.6 Z$_{\rm \odot}$, while further away along
jet axis the contrast is 0.1-0.2. This kind of feature is supported by recent
deep {\it Chandra} observations done by \citet{kir09}. 
They produced Fe-maps showing
the enrichment of gas along the jets of Hydra A, a quite massive galaxy cluster.
It is striking that our simulated maps resemble these observations also in a quantitative way.
Other authors, Doria et al. (in preparation), found a similar 
behaviour in a different cluster dominated by AGN activity (RBS797).

At 3.5 Gyr (second panel) we are in a period of relatively quiescence and so
the now dominating (AGN induced) turbulence and vorticity 
promote the iron diffusion. At the centre, the new iron cloud associated to the cD galaxy 
is clearly detached from the older ejected material, the latter covering now a more uniform area.
The iron diffusion is more evident at 6.5 Gyr (third panel), again in a moment of AGN quiet, where
the iron is very diffuse, with a low enhancement of 0.05-0.1 Z$_{\rm \odot}$.

In the end we can affirm that in a single cycle the iron abundance
passes from a phase of high asymmetry along jet axis, when the outflow
has recently turned on, to a phase of turbulence and mixing, in which
spherical symmetry is almost restored. The iron gets spread within few
100 kpc, just as observed: without the influence of an AGN such material
would be indeed confined near the effective radius of the central galaxy.

\subsection{Hydrostatic equilibrium}

In this Section we investigate the effect of the perturbations
generated by the AGN outflows on the mass determination using
X-ray observations (through Eq. (\ref{EH})).
A standard method to estimate galaxy cluster mass profiles is
to assume spherical symmetry and estimate the gravitational
mass using the hydrostatic equilibrium equation (e.g. \citealt{buo07,vik06} 
for two recent compilations). 
When the flow velocity is much lower than the sound speed,
the hydrostatic assumption is fully justified.

It is well known that in real clusters turbulence or large
scale motion can systematically bias the mass estimate 
by $\lta 20$\% (\citealt{ras06, nag07}; \citealt{mah08,piv08}).
Therefore, it is interesting to investigate how much the 
flow perturbations generated by the AGN feedback affect
the mass measure.

Here we quantify the discrepancy between the estimated
gravitational mass and the real one for a model adopting feedback scheme A
and one assuming the more quiet feedback Bondi.
We have calculated the azimuthally averaged density and emission weighted 
temperature profiles in rings of
progressively increasing width, from 10 kpc in the centre
up to 100 kpc at large radii.
These profiles, inserted in the hydrostatic equilibrium equation
give the estimated mass profile. This procedure is not perfectly
accurate: we should have `observed' our simulations with specific
softwares like X-MAS (\citealt{gar04,ras08})
or XIM (\citealt{hei09}) to properly compute the averaged profiles. 
However, it is not the purpose of this
paper to thoroughly investigate this topic and our approximate analysis
is sufficient to make our point.

The calculated mass profiles at various times, as well as the exact
mass profile, are shown in Fig. 15 for model A3 (top panel) 
and BONDI2 (bottom panel).

In run A3 the powerful outflows, although not able to significantly perturb
the thermal state of the cluster (which always preserves the cool core appearance),
are effective in disturbing the quasi-hydrostatic equilibrium present in the
pure cooling flow model. This results in a typical error in the mass determination
of a factor 2-3. Most likely the estimated mass is lower than the real one,
and the discrepancy is larger in the region $r\lta 100$ kpc, where the feedback
affect the ICM the most.

In model BONDI2, instead, the outflows are 3-4 orders of magnitude less powerful
and, despite the fact that they are continuously generated, they seem rather innocuous for the
dynamical state of the ICM and the hydrostatic equilibrium approximation is safe.
The computed mass profile agrees very well with the real one, with slight discrepancy
only visible for $r\lta 20$ kpc, a region inadequately resolved in our simulations.
Of course, the weak, continuous jets present in this run are incapable to generate
cavities (see Section 3.5), hence it is likely that real clusters undergo at least few
stronger AGN outbursts, after which the dynamics of the ICM would be similar to that
described for model A3.

In summary, we expect that for clusters where X-ray cavities and/or shocks
are present the total mass estimated through the assumption of hydrostatic equilibrium
might be in error by a factor of $\sim 1.5$, occasionally the error could get
as large as a factor of 2-3. Conversely, with gentler continuous outflows
generated by the relatively accretion rate predicted by Bondi theory, the estimated
and the real mass are in excellent agreement, with errors always below $8$\%
(see also \citealt{gum10}).

\begin{figure}
\centering
\includegraphics[width=57mm]{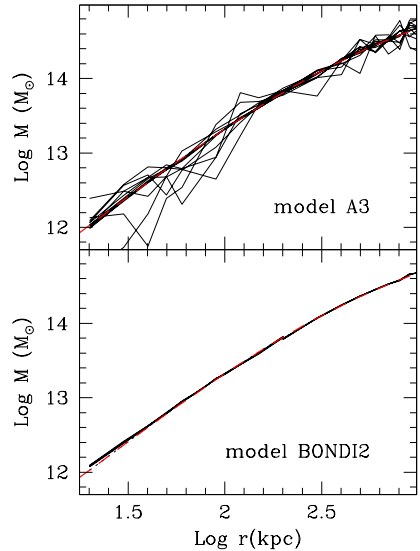}
\caption{Mass profiles for model A3 (top panel) and BONDI2
(bottom panel), calculated using the hydrostatic equilibrium
equation at times separated by 0.5 Gyr (model A3) and by 1 Gyr
(model BONDI2),
and the exact mass profile (red dashed line).} \label{fig:hydroeq} 
\end{figure}

\section[]{Discussion}

In this paper we have presented several moderate resolution
simulations of the interaction between AGN outflows and the ICM
in massive galaxy clusters. The purpose of these calculations
was to investigate if a purely mechanical AGN feedback of this kind is able to solve
the so called {\it cooling flow problem}: the dearth of cooling
gas in cool core clusters with short central cooling times.

The necessity of covering a large range of
spatial (kpc up to 5 Mpc) and temporal scales (fraction of Myr up
to 7 Gyr) limited our resolution of the very inner accretion region.
Hence we had to link the feedback to some large scale mean quantity, like $\Delta M_{\rm cold}$
or $\dot{M}_{\rm Bondi}$, with the obvious result of some discrepancies with observables (like bubbles).
In any case gas accretion onto the SMBH and subsequent jet-outflow ignition is
still obscure, in particular the amount of ICM entrained and shocked.
Furthermore, we do not have a long term (Gyr) evolution of
theoretical models for accretion, due to computational limitations.
In the end our simplified scheme seems a good and efficient method to test AGN feedback 
on galaxy cluster scales. 

\subsection{Comparison with other works}

In the field of simulated AGN feedback most of the work done in the past (\citealt{brk02,brm03,rus04,dal04,bru05,sij07,brs09,mab08a,mab08b,mat09}) 
has focused on the 
creation of `artificial' bubbles off-centre.
This scenario is motivated by the fact that relativistic radio jets (distinct from our outflows)
do not entrain much gas and just thermalize the ICM at the hotspot, generating a cavity.

The difference with our study is that we employ a momentum-driven heating, instead of
a pressure-driven one. As we have seen, purely kinetic feedback changes the whole dynamics.
Intermittent bubbles are naturally created, but this shocked and expanding gas is just one element of the heating process\footnote{
usually most of the mechanical energy is not used to form the cavity (see Sec. 4)}
: the bow shock, the cocoon with entrained gas, the mixing through turbulence and vorticity are all fundamental
gears of the machine, everyone dominant at different phases of evolution.
Moreover, artificial (hydrodynamical) bubbles
becomes unstable rapidly; on the contrary some studies (\citealt{ste07})
indicate that jet-inflated ones may stay intact for over 40 Myr, thanks
to the vortex formed inside. It is probable that a combination of the two
mechanisms is required (outflow plus artificial buoyant bubble), even if it 
still unclear which of the two dominates. 

It is difficult to compare our work with other kinetic outflows simulations
(e.g. \citealt{rus04,omm04,zan05,ste07}), because most of them do not
implement radiative cooling and an 
initial cool core state. Additionally, they last for few hundreds Myr and 
are not suited to study the `cooling flow problem', without the long term evolution.
 
The only investigations done in this direction are those of BM06 and \citealt{cat07}. 
In the first one, as many previous studies, the jet velocity and power were set by-hand
following observational estimates. We have seen (feedback scheme B), that in imposing such predetermined conditions, 
the mechanical feedback is not linked to the cooled mass or lower gas entropy (in quantity and time). 
The results are models with temperature and density profiles similar to a pure cooling flow
or with high cooling rates.

In \citet{cat07} the injection scheme is self-regulated by Bondi accretion (for 12 Gyr), 
but after a weak burst of $\sim10^{45}$
erg s$^{-1}$ the profiles do not present the cool core appearance (flat temperature).
Also our successful Bondi model is able to prevent the cooling catastrophe.
The difference is that our profiles present ripples and signature of weak shocks.
It is again difficult to make a full comparison, because a large fraction
of their injected energy is thermal, not only kinetic.

\citet{dub10} expanded the previous model in a cosmological resimulation.
Even if they focused on the BH growth history, they found Bondi outflows can prevent
very peaked density profiles, despite the presence of negative temperature
gradient at various times. Here the main difference with our simulations (and thus 
some results) is probably the cosmological context. 
We calculate a fully relaxed cluster, while they analyse a 1:1 merging event. 
In any case, we underline
the fact that all these results give one clear indication: AGN outflows are a key
component of the feedback in galaxy clusters.
 
The last comparison is the one with \citet{ver06}. 
They implemented a self-regulated mechanism similar to our boundary injection scheme.
Their simulations present very different results from our (and Cattaneo et al.) analysis.
After few hundreds Myr (with every AGN model) the cooling becomes catastrophic: $>2000$ $\msun$ yr$^{-1}$.
The problem of the creation of an unidirectional channel and subsequent
deposition of energy is a difficulty found also in our outflow models, but
partially resolved in the long term (see Section 6.3.1).

In the next Sections we will deeply discuss the merits and
flaws, previously introduced, of our simulated feedback models.

\subsection{Cold-regulated explosive feedback (A)}
 
In all our study we covered a wide `zoology' of triggering mechanisms,
varying the mechanical efficiency and injection method.

In the first series of models we linked the power of the AGN
outflows to the cooling rate (feedback scheme A, Sections 3.1), 
the latter
being calculated as $\Delta M_{\rm cool} / \Delta t$, where
$\Delta M_{\rm cool}$ is the mass of gas cooled within the region
$r\lta 10$ kpc in a single timestep $\Delta t$.
It is clear that the energy of a single jet event depends
on numerical resolution through the Courant-Friedrichs-Lewy (CFL)
stability condition. Lower resolution calculations, with larger
$\Delta t$ and thus larger $\Delta M_{\rm cool}$, 
will generate more powerful outburst events.

Although in the models described in Section 3.1 the
total kinetic energy injected at a given time
rests on the integrated cooled gas mass (which is only weakly
dependent on the numerical resolution),
and can in principle vary among
different models (i.e. different $\epsilon$),
we find, somewhat surprisingly, that it is insensitive on the 
efficiency $\epsilon$ and always about $2-3 \times 10^{62}$ erg,
a value comparable to the total energy radiated away.
Evidently, the self-regulation mechanism assumed is very effective.

The first astrophysically important result of these calculations
is that the effect of the feedback crucially depends on the
modality of the energy injection, not only on the
total amount of energy.
We have found that when the efficiency is low and the 
outflows are frequent and relatively weak (models A1 and A2), 
the gas cooling 
proceeds at a rate much higher than the limit allowed by observations.

Conversely, when $\epsilon \gta 0.01$ (models A5 and A6) the cooling rate
is reduced to a value fully consistent with the observational
constraints. However the ICM in the core is heated too much and the
temperature and density profiles do not resemble those of a
typical cool core cluster.

Only models A with efficiency in the range
$5\times 10^{-3} \lta \epsilon \lta 0.01$ are
able to reduce the cooling rate to acceptable values
preserving at the same time a central positive temperature gradient.
Moreover, these models generate cavities and shocks, another important
requirement that a plausible heating scenario must satisfy.
Notice that also observational data (\citealt{meh08,laf10})
suggest an average mechanical efficiency of AGN around $5\times 10^{-3}$.

As pointed out before, other more ad hoc types of feedback (not directly linked to
cooled gas), like intermittent (B) or
continuous with fixed velocity (C), present a cooling flow appearance, 
in the former case,
or a too high total energy injection, in the latter. These models are not totally
disastrous and show some good features (as found in BM06), but in the long term
they are not satisfactory.

\subsubsection{Riddles}

Adopting a severe critical point of view, successful A models may present two
difficulties. First, gas often cools and accretes onto the black hole
at a rate far above the Eddington limit
(section 3.4), a fact which make them unpalatable (\citealt{kin09}), but
not impossible (especially at higher redshift, see for example \citealt{szu04,gho10}).
It is known that BHs grow mainly through the `QSO phase',
that is through radiative accretion at early times (\citealt{hnh06}). 
Therefore our accretion rates seem overestimated, considering that
the BH mass should not increase much in the entire simulation time.

Remember, however, that our resolution does not permit to track
the exact amount of accreted material on sub-pc scale. Hence it is probable that only
a small part of $\Delta M_{\rm cool}$ falls onto black hole, and the other
shall be just entrained or ejected by the outflow. It may
be possible that the real efficiency could reach higher values (instead
of $10^{-2}-10^{-3}$ adopted in our successful simulations).

For example, follow this consideration and let us define the inflow rate
as the sum of the real BH accretion rate and the gas outflow rate: 
$\dot{M}_{\rm in} \equiv \dot{M}_{\rm acc} + \dot{M}_{\rm out}$. 
If $\eta \equiv \dot{M}_{\rm out}/\dot{M}_{\rm acc}$, then the jet power
$P_{\rm jet} \equiv \epsilon_{\rm real} \dot{M}_{\rm acc} c^2 = \epsilon_{\rm real} \dot{M}_{\rm in} c^2 / (1+\eta)$.
Thus we can say that the mechanical efficiency adopted 
in our simulations could be $\epsilon = \epsilon_{\rm real}/ (1+\eta)$.
A mean $\eta$ can be retrieved as $\Delta M_{\rm in}/\Delta M_{\rm acc} -1$.
Now, assuming a small BH mass variation in 7 Gyr of
0.1 $M_{\rm BH}$ (that is $\Delta M_{\rm acc}\simeq 3\times10^8$ $\msun$), and noting
that for model A3 $\Delta M_{\rm in}=\Delta M_{\rm cold}\simeq 3\times10^9$ $\msun$, then clearly $\eta\sim9$.
With this in mind, the real efficiency may be  
$\epsilon_{\rm real} = (1+\eta) \epsilon \sim 10\epsilon=5\times10^{-2}$.
Not only, we can also estimate $\Delta M_{\rm out}\simeq 2.7\times10^9$ $\msun$ and,
with a duty cycle of 10\%
\footnote{Thus a single jet event has $\Delta M_{\rm out}\sim 10^7 - 10^8$ $\msun$}, the mean outflow rate results in a few $\msun$ yr$^{-1}$.
From here we can check the entrainment (or `mass loading'): 
$\lambda_{\rm en} = \dot{M}_{\rm act}/\dot{M}_{\rm out}$, where 
$\dot{M}_{\rm act}$ is the outflow rate in the active region. Taking a mean value of
$10^3$ $\msun$ yr$^{-1}$, the entrainment is $\lambda_{\rm en}\sim10^2-10^3$.

Regarding entrainment, in this first series of 3D simulations we just
focused on the global consequences of massive outflows on the thermal and
dynamical ICM evolution on large scales. We assume that the outflow is 
momentum-driven with negligible thermal energy. The detailed process of
generating the entrainment is here not addressed. One possible explanation
is given by \citealt{sok08}, whose model forms slow massive wide jets.

Summarizing, in order to have a small BH growth, in one of our best models,
we have to assume massive outflows with a rate $\sim9$ times the accretion one.
The ratio $\eta$ is not well constrained, but for example \citet{moe09}
give values around 10. 
Nowadays nobody knows how much gas is really loaded by the original jet,
but $\sim10^2-10^3$ seems reasonable (\citealt{cat07} adopted 100 for example). 
This could be in the future another observational constraint for our numerical models.

We conclude from all this reasoning that the high `accretion' rates in our successful simulations can
be easily reduced, introducing an higher (real) efficiency and  
considering that only a fraction of $\Delta M_{\rm cold}$ falls onto BH, while the other part
is ejected and entrained.

The second flaw in models A may be that the process of cavity formation heats
the surrounding ICM too much (Section 5.1): soon after jet ignition the temperature
inside the bubble rises to high value ($\sim 10^8$ K), and the ICM around is highly shocked.
On the other side many observations (\citealt{fab00,mac00,bla01,nul02,bla03,fab03,fab06}; MN07) suggest that
the bubble inflation should be gentler, in order to produce rims of gas cooler
than the ambient medium. 
Again our resolution does not permit an accurate and detailed study of this and other
kind of features (e.g. cold filaments). Certainly, a direct cause of the violent behaviour is that
the injection energy in a single timestep is too much. This will
probably be avoided with higher spatial resolution, because of smaller $\Delta t$
and therefore smaller $\Delta  M_{\rm cold}$ and blowing instant power. 
Unfortunately with current
computational resources we were limited to a very short evolution. In
any case we plan to develop such a study in a future work.

We can also argue that observations may be seeing bubbles at a late time, well
after their generation, and this could explain lower temperature and SB jumps. In fact
of the three cavities analysed in Fig. 13, just the newly born one present sharp
contrasts, while the other are older and hence feebler. Moreover, we can not exclude
that in the entire simulation some bubbles are created in a gentler way by weaker bursts
($\sim10^{46}$ erg s$^{-1}$).

\subsection{Hot-regulated gentle feedback (Bondi)}

To circumvent some of the aforementioned difficulties encountered with
feedback A method, we changed completely direction, calculating 
models where the accretion
rate was set to the Bondi rate. Contrary to models A, this generates a 
continuous feedback of moderate power (typically few $\times 10^{44}$
erg s$^{-1}$). In order to be efficient, such non explosive jet power
must be sufficiently steady in order to reduce cooling flow. As pointed out
in Section 2.2, our resolution is far above Bondi radius ($\sim50$ pc), thus we have
to rely again upon mean large scale quantities. In this case we must stay
as much possible close to the BH, avoiding the simplest case of a few cells,
because of numerical fluctuations. When the radius of ($\rho$, $c_{\rm s}$) averaging
is fairly small, $\sim 5$ kpc, the power is steady and the model successful:
cooling rate are low, profiles follow observations, and
energies are contained.

On the other side, with $r_{\rm av}\sim 10$ kpc, we revert again to a spasmodic
feedback very sensitive to central cold mass. Hence this Bondi model
recalls low efficiency A1 model. This behaviour is also emphasized by
the fact that here we are injecting mechanical energy directly in
the domain cells, and not through a boundary (like BONDI2). In a short
bursting event like models A this difference is not relevant, while
with a continuous outflow, pushing the gas from below and not changing
suddenly internal flow variables, seem to be a more efficient way to heat
the inflowing central medium. 

This illustrates the (unfortunate) sensitivity of the
simulations on some numerical detail. The cause is that
any kind of AGN feedback simulation with large spatial and temporal 
scale integration must require a few quantities
set by-hand. In general a fully self-consistent simulation 
is rather utopian, for now. In our case the ignorance of detailed 
AGN accretion and jet-outflow physics, plus today computational
resources, limits us to simplified numerical feedback models.
Nevertheless, after trying many possible models and different 
parameters, they let us study very well the consequences
of this assumptions on scales of interest.        

\subsubsection{Riddles}

A flaw we found in successful Bondi models is that
they do not naturally generate X-ray cavities. Instead, the continuous 
outflows carve a narrow and long tunnel along the $z$-axis. However,
we have seen that continuous jets, with ram pressure
almost equal to gas thermal pressure, 
are highly disrupted by turbulence, especially in central regions.
Hence, the fragmentation of a feeble jet may produce 
generations of `gentle' detached bubbles, which rise buoyantly, even in continuous feedback models.

We propose that any kind of strong turbulence
can indeed disrupt a moderate power jet. For example, \citet{mor10} and \citet{dub10} tried to start 
with a cluster atmosphere in non hydrostatic equilibrium, i.e. following a real
cosmological evolution (local density fluctuations, merging, etc.), and they found
a fragmentation of the AGN jets, helping the deposition of energy in the inner core.

The creation of an unidirectional channel and subsequent energy deposition
at too large radii was pointed out by \citet{ver06}.
However, if we perform long term evolutions, we have showed that the
above mentioned turbulence and vorticity
promotes mixing in the central active region, replenishing again the channel. Even in BONDI2 model,
where a small narrow channel always stays open, instabilities and turbulence clearly heat the gas at
the base of the jet, letting to cool only a moderate amount of gas in the near equatorial region.

\subsection{Best models confrontation}

It is interesting to find other important differences (or similarities) in successful
models A and Bondi. The latter displays low velocities of the order $6-7\times10^3$ km s$^{-1}$,
a value consistent with line-absorption observations (see \citealt{cre03} for a review,
and other references in Section 1). Explosive models A tend to have also higher velocities, in a few 
events reaching $\sim 10^5$ km s$^{-1}$, more similar to a fast jet than an outflow wind.
Both models have reasonable final injected energies,
below a theoretical total energy of a BH with mass $10^9$, $\sim2\times10^{62}$ erg.

Apart from cavities and shocks (previously discussed), we produce other significant predictions,
which are (or will be) comparable to X-ray observations. Iron abundance maps
play a relevant role in tracing the outflow activity. The metals, produced
mainly by SNIa in the cD elliptical galaxy, are easily transported along jet-axis
up to 150-200 kpc, creating an unmistakable asymmetry, when the AGN is very active. 
The observations by
\citet{kir09} and Doria et al. (prep.) confirm this behaviour. In the period of 
quiescence turbulence and bulk motion will dominate the scene, smoothing
and almost restoring the homogeneity. This could be another constraint in choosing 
between Bondi or not: in fact a continuous jet will tend to show often this
marked asymmetry.

Finally, another striking diversity is the clean departure
from hydrostatic equilibrium in models A, due to its more explosive nature. On
the contrary, in Bondi type, mean fluctuations are very contained with errors below 8\%.
The consequence is, in the first case, a less precise mass determination using Eq.
(\ref{EH}), as also found by other observational works (\citealt{nag07,ras06}).   

We conclude pointing out that the gaps of models A, can be replenished by the features of 
Bondi feedback,
and viceversa (especially for cavities). Thus, an intriguing solution may be a `dual model', in which 
few energetic AGN outbursts (like models A3-4), perhaps triggered by accretion rates close to the Eddington limit,
are superimposed (or alternated) to the weak activity induced by Bondi prescription. The first explosive mode will create large spheroidal bubbles, as those observed in real clusters, while the second quasi-steady outflow will be the real sustaining pillar of the heating machine. This scenario would require a physical explanation of the alternation mechanism or 
why the two types of outflow are diversified.

\section[]{Conclusions}
Overall we found that
subrelativistic AGN outflows, produced by two types of self-regulated feedback,
are able to quench cooling flow for at least 7 Gyr and, at the same time, preserve
the cool core appearance:\\
\begin{description}
\item (a) self-regulated feedback based on 
the instantaneous $\Delta M_{\rm cool}$ accreted, with mechanical
efficiencies between $\sim5\times10^{-3}-10^{-2}$;\\
\item (b) Bondi triggered feedback with $\epsilon$ around $0.1$, based on 
an almost continuous and steady outflow, generated by hot gas accretion.\\
\end{description}

Both best models, (a) and (b), present primary merits concordant with X-ray observations:
\begin{description}
\item (1a,\,b) cooling rate is reduced at least under 5\% of the pure CF model;\\
\item (2a,\,b) (mass-weighted) $T(r)$ and $n(r)$ profiles oscillate \\
near the observed ones;\\
\item (3a,\,b) total injected energy is always below $2\times10^{62}$ erg, under the limit of $E_{\rm BH}$;\\
\item (4a,\,b) mean $v_{\rm jet}$: from $\sim5\times10^{3}$ to a few $10^4$ km s$^{-1}$.
\end{description}

Intermittent (B) or continuous scheme with fixed velocity (C) are not
consistent in the long term with some of these points, and hence rejected.

In addition to these constraints we wanted to test further the best models.
We warn that
the following are supplementary predictions, but not main goals of our study:
due to the need of a long term evolution,
our simulations are limited in resolution ($\sim$ kpc), and fine details could
be altered or just missed by our numerical implementation.  
Marked differences between the two best models are:
\begin{description}
\item (5a) super-Eddington $\dot{M}_{\rm acc}$: high power ($\sim10^{48}$ erg s$^{-1}$);\\
\item (5b) sub-Eddington $\dot{M}_{\rm acc}$: low power ($\sim10^{44}$ erg s$^{-1}$);\\
\item (6a) cavities with high internal energy and shocked rims;\\
\item (6b) absence of a duty cycle and real inflated bubbles;\\
\item (7a) asymmetrical transport of metals (SNIa) 
along jet-axis up to 200 kpc; subsequent gradient lengthened by turbulence;\\
\item (7b) almost always asymmetrical iron maps;\\
\item (8a) large deviation from hydrostatic equilibrium state (high turbulence);\\
\item (8b) deviation from HE mass determination below 8\% (moderate turbulence).
\end{description} 

Points (6) (caused by (5)) seem not to follow entirely observations (even if
we need larger samples of clusters with AGN activity to reconstruct global history,
and certainly higher resolved SB maps to study cavities). While the two
models appear at first sight antithetical, it might be possible that they alternate
during evolution, or that high power outbursts are just superimposed
to a low Bondi feedback. 

Despite these riddles, purely mechanical AGN outflows promise to be good candidates,
to uphold the wild and frenetic `dance' between heating and cooling.

\section*{Acknowledgments}
The software used in this work was in part developed by the DOE
supported ASCI/Alliances Centre for Thermonuclear Flashes at the
University of Chicago.
We acknowledge the CINECA-INAF 2008-2010
agreement for the availability of high performance computing resources,
and the support by MIUR grant PRIN 2007C53X.
We would like also to thank Noam Soker and Fulai Guo for the useful discussion.

\bsp

\label{lastpage}

\end{document}